\documentclass[11pt,prd,onecolumn,amsmath,amssymb,aps,floats,floatfix,nofootinbib]{revtex4-2}
\usepackage[colorlinks=true,urlcolor=blue,anchorcolor=blue,citecolor=blue,filecolor=blue,linkcolor=blue,menucolor=blue,linktocpage=true]{hyperref} 

\usepackage[inline]{enumitem}
\usepackage[multidot]{grffile}  
\usepackage{dcolumn}
\usepackage{bm}
\usepackage{amsmath}
\usepackage{amsfonts}
\usepackage{amssymb}
\usepackage{color}
\usepackage[table]{xcolor}
\usepackage{float}
\usepackage{latexsym}
\usepackage{slashed} 
\usepackage{pstricks}
\usepackage{indentfirst}
\usepackage{mathrsfs}
\usepackage{multirow}
\usepackage{epsfig,psfrag}
\usepackage{graphicx}
\usepackage{enumitem}
\usepackage{geometry,amssymb,yfonts}
\usepackage{yhmath}
\usepackage{anysize}
\usepackage{subfigure}
\usepackage{mathtools}
\usepackage{setspace} 
\usepackage[utf8]{inputenc} 
\usepackage[scientific-notation=true]{siunitx} 

\usepackage{url}
\usepackage{makecell}

\graphicspath{{fig/}}

\allowdisplaybreaks 

\newcommand{\Uone}{\mathrm{U}(1)}
\newcommand{\UoneX}{\mathrm{U}(1)_\mathrm{X}}
\newcommand{\UoneY}{\mathrm{U}(1)_\mathrm{Y}}

\newcommand{\SUtwoL}{\mathrm{SU}(2)_\mathrm{L}}
\newcommand{\SUthreeC}{\mathrm{SU}(3)_\mathrm{C}}

\begin{document}

\title{Gravitational waves from cosmic strings associated with pseudo-Nambu-Goldstone dark matter}

\author{Ze-Yu Qiu}
\author{Zhao-Huan Yu}\email[Corresponding author. ]{yuzhaoh5@mail.sysu.edu.cn}
\affiliation{School of Physics, Sun Yat-Sen University, Guangzhou 510275, China}

\begin{abstract}
We study stochastic gravitational waves from cosmic strings generated in an ultraviolet-complete model for pseudo-Nambu-Goldstone dark matter with a hidden $\mathrm{U(1)}$ gauge symmetry. The dark matter candidate in this model can naturally evade direct detection bounds and easily satisfy other phenomenological constraints. The bound on the dark matter lifetime implies an ultraviolet scale higher than $10^9~\mathrm{GeV}$. The spontaneous $\mathrm{U(1)}$ symmetry breaking at such a high scale would induce cosmic strings with high tension, resulting in a stochastic gravitational wave background with a high energy density. We investigate the constraints from current gravitational wave experiments as well as the future sensitivity. We find that most viable parameter points can be well studied in future gravitational wave experiments.
\end{abstract}

\maketitle
\tableofcontents

\clearpage

\section{Introduction}

Null experimental results from the direct detection of dark matter (DM) have put strong constraints on the DM-nucleon scattering cross section~\cite{PandaX-4T:2021bab, Aalbers:2022fxq, XENON:2023sxq}, implying that DM particles might not have a weak interaction strength, which is required by the conventional freeze-out mechanism of DM production~\cite{Bertone:2004pz, Feng:2010gw, Young:2016ala}.
Nevertheless, it is possible to suppress DM-nucleon scattering but leave the DM annihilation cross section at the freeze-out epoch unaffected.
This can be gracefully realized by assuming that the DM particle is a pseudo-Nambu-Goldstone boson (pNGB) whose scattering off nucleons is extremely suppressed by low momenta~\cite{Gross:2017dan, Azevedo:2018exj, Ishiwata:2018sdi, Huitu:2018gbc, Alanne:2018zjm, Kannike:2019wsn, Karamitros:2019ewv, Cline:2019okt, Jiang:2019soj, Ruhdorfer:2019utl, Arina:2019tib, Abe:2020iph, Okada:2020zxo, Glaus:2020ihj, Abe:2020ldj, Zhang:2021alu, Abe:2021byq, Okada:2021qmi, Abe:2021jcz, Abe:2021vat, Biekotter:2021ovi, Zeng:2021moz, Cai:2021evx, Mohapatra:2021ozu, Darvishi:2022wnd, Abe:2022mlc, Biekotter:2022bxp, Liu:2022evb, Otsuka:2022zdy, Gola:2022nkg, Jiang:2023xdf, Mohapatra:2023aei}, evading the constraints from direct detection experiments.

Such a pNGB DM framework typically requires a proper ultraviolet (UV) completion with an extra gauged $\Uone$ symmetry~\cite{Gross:2017dan, Abe:2020iph, Okada:2020zxo, Liu:2022evb}.
The experimental bound on the DM lifetime demands that the UV scale where this gauge symmetry breaks down could be higher than $\mathcal{O}(10^{10})~\si{GeV}$~\cite{Abe:2020iph, Okada:2020zxo, Liu:2022evb}.
The spontaneous breaking of the $\Uone$ gauge symmetry in the early universe may results in cosmic strings~\cite{Nielsen:1973cs, Kibble:1976sj}, whose tension, \textit{i.e.}, the energy per unit length, would be rather large because of the high UV scale.
The network of cosmic strings are expected to radiate 
gravitational waves (GWs)~\cite{Vilenkin:1981bx, Hogan:1984is}, which would form a stochastic background remaining as a relic in the present universe.

The stochastic GW background (SGWB) from cosmic strings covers an extremely broad range of GW frequencies~\cite{Vilenkin:2000jqa}.
Therefore, such a background is an interesting target for various types of GW experiments in different frequency bands, including pulsar timing arrays (PTAs) in $10^{-9}\text{--}10^{-7}~\si{Hz}$, ground interferometers in $10\text{--}10^{3}~\si{Hz}$, and future space interferometers such as LISA~\cite{LISA:2017pwj}, TianQin~\cite{TianQin:2020hid, Cheng:2022vct}, and Taiji~\cite{Hu:2017mde} in $10^{-4}\text{--}10^{-1}~\si{Hz}$.
As a result, searching for an SGWB provides a unique way to test many new physics theories beyond the standard model (SM) that would induce cosmic strings in the early universe~\cite{Buchmuller:2013lra, Dror:2019syi, Gouttenoire:2019kij, Gouttenoire:2019rtn, Buchmuller:2019gfy, Blasi:2020wpy, King:2020hyd, ZhouRuiYu:2020bbs, Fornal:2020esl, Chigusa:2020rks, Lazarides:2021uxv, King:2021gmj, Samanta:2020cdk, Masoud:2021prr, Bian:2021vmi, Samanta:2021zzk, Dunsky:2021tih, Samanta:2021mdm, Cai:2021dgx, Ahmed:2022rwy, Afzal:2022vjx, Borah:2022byb, Lazarides:2022jgr, Yamada:2022imq, Borah:2022vsu, Maji:2022jzu, Fu:2022lrn, Lazarides:2022ezc, Hindmarsh:2022awe, Borah:2022iym, Saad:2022mzu}.

In this work, we will study the stochastic GWs radiated by cosmic strings originating from the UV-complete theory of pNGB DM.
As mentioned above, the required high UV scale leads to cosmic strings with a rather high tension, which results in a high energy density of stochastic GWs.
Thus, GW experiments could be more sensitive to this theory than other types of experiments, which would typically lose sensitivity for high UV scales.
More specifically, we will discuss a simple model with a hidden $\Uone$ gauge symmetry~\cite{Liu:2022evb} as an illuminating example and explore the constraints from previous GW experiments and the sensitivities of future GW experiments.

The remainder of this paper is organized as follows.
In Section~\ref{sec:model}, we briefly introduce the UV-complete model for pNGB DM with a hidden $\Uone$ gauge symmetry.
In Section~\ref{sec:phen}, we study the phenomenological constraints and perform a random scan in the parameter space.
In Section~\ref{sec:CS}, we discuss the SGWB arising from cosmic strings, which are induced by the spontaneous breaking of the $\Uone$ gauge symmetry.
In Section~\ref{sec:GW}, we investigate the constraints from current GW experiments and estimate the sensitivity in future GW experiments.
In Section~\ref{sec:sum}, we summarize the paper.

\section{pNGB DM model}
\label{sec:model}

In this section, we briefly discuss the UV-complete model of pNGB DM that extends the SM with a hidden $\UoneX$ gauge symmetry.
Details can be found in Ref.~\cite{Liu:2022evb}.
``Hidden'' means that all the SM fields do not have $\UoneX$ charges.
Two complex scalar fields $S$ and $\Phi$ carrying $\UoneX$ charges $1$ and $2$ are introduced, and they are singlets under the SM gauge group $\SUthreeC \times \SUtwoL \times \UoneY$.
We denote the SM Higgs field as $H$.
The related Lagrangian is
\begin{eqnarray}\label{eq:Lag}
\mathcal{L} &\supset&  -\frac{1}{4}W^{a,\mu\nu}W^a_{\mu\nu} -\frac{1}{4}B^{\mu\nu}B_{\mu\nu} -\frac{1}{4}X^{\mu\nu}X_{\mu\nu} - \frac{s_{\varepsilon}}{2}B^{\mu\nu}X_{\mu\nu}
\nonumber\\
&& +(D^{\mu}H)^{\dag}(D_{\mu}H)+(D^{\mu}S)^{\dag}(D_{\mu}S)+(D^{\mu}\Phi)^{\dag}(D_{\mu}\Phi)
\nonumber\\
&&
+\mu^{2}_H |H|^{2}
+\mu_{S}^{2}|S|^{2}
+\mu_{\Phi}^{2}|\Phi|^{2}
-\frac{\lambda_{H}}{2}|H|^{4} -\frac{\lambda_{S}}{2}|S|^{4}
-\frac{\lambda_{\Phi}}{2}|\Phi|^{4}
\nonumber\\
&&-\lambda_{HS}|H|^{2}|S|^{2}
-\lambda_{H\Phi}|H|^{2}|\Phi|^{2}
-\lambda_{S\Phi}|S|^{2}|\Phi|^{2}
+\frac{\mu_{S\Phi}}{\sqrt{2}}(\Phi^{\dag}S^{2}+\Phi S^{\dag 2}),
\end{eqnarray}
where $W^a_{\mu\nu} = \partial_{\mu}W^a_\nu - \partial_{\nu}W^a_\mu + g\varepsilon^{abc}W^b_\mu W^c_\nu$, $B_{\mu\nu} \equiv \partial_\mu B_\nu - \partial_\nu B_\mu$, and $X_{\mu\nu} \equiv \partial_\mu X_\nu - \partial_\nu X_\mu$ are the field strengths of the $\SUtwoL$, $\UoneY$, and $\UoneX$ gauge fields $W^{a,\mu}$, $B^\mu$, and $X^\mu$.
The parameter $s_\varepsilon \equiv \sin\varepsilon$ induces the kinetic mixing between $B^\mu$ and $X^\mu$.
The covariant derivatives are given by $D_{\mu}H = (\partial_{\mu}-ig' B_{\mu}/2 - \mathrm{i}g W_\mu^a \sigma^a/2)H$,
$D_{\mu}S = (\partial_{\mu}-\mathrm{i} g_{X}X_{\mu})S$,
and $D_{\mu}\Phi = (\partial_{\mu}- 2\mathrm{i} g_{X}X_{\mu})\Phi$,
with $g_X$ denoting the $\UoneX$ gauge coupling.

For realizing the pNGB DM framework, $S$ and $\Phi$ gains nonzero vacuum expectation values (VEVs) $v_S$ and $v_\Phi$, which satisfy $v_S\ll v_{\Phi}$.
Thus, $v_\Phi$ is a UV scale below which the $\UoneX$ gauge symmetry spontaneously broken to a $\UoneX$ global symmetry, which is approximate, because the $\mu_{S\Phi}$ term softly breaks it.
$v_S$ is a lower scale where the $\UoneX$ global symmetry is spontaneously broken, leading to a pNGB with a mass arising from $\mu_{S\Phi}$.
Such a pNGB is a DM candidate whose scattering off nucleons is extremely suppressed~\cite{Gross:2017dan}.

We decompose the scalar fields as
\begin{eqnarray}
H=\frac{1}{\sqrt{2}}\begin{pmatrix}
0 \\
v+h
\end{pmatrix},\quad
S=\frac{1}{\sqrt{2}}(v_{S}+s+\mathrm{i}\eta_{S}),\quad
\Phi=\frac{1}{\sqrt{2}}(v_{\Phi}+\phi+\mathrm{i}\eta_{\Phi}),
\end{eqnarray}
where the SM Higgs VEV is $v = 246.22~\si{GeV}$. 
There are mass mixing terms among the $CP$-even scalars $(h, s, \phi)$ and between the $CP$-odd scalars $(\eta_S, \eta_\Phi)$.
After diagonalizing the mass-squared matrices, we obtain the physical scalar mass terms
\begin{equation}
\mathcal{L}_{\mathrm{mass}} \supset -\frac{1}{2} \sum\limits_{i = 1}^3 m_{h_i}^2 h_i^2 - \frac{1}{2} m_\chi^2 \chi^2.
\end{equation}
The Higgs bosons $h_1$, $h_2$, and $h_3$ are linear combinations of $h$, $s$, and $\phi$.
We further require that $h_1$ is the SM-like Higgs boson, whose mass is measured to be $m_{h_1}=125.25 \pm 0.17~\si{GeV}$~\cite{ParticleDataGroup:2022pth}, and that $h_2$ and $h_3$ are the $s$-like and $\phi$-like exotic Higgs bosons, respectively.
$\chi$ is a linear combination of $\eta_S$ and $\eta_\Phi$ with a mass given by
\begin{eqnarray}
m_{\chi}= \sqrt{\frac{\mu_{S\Phi}(v_{S}^{2}+4v_{\Phi}^{2})}{2v_{\Phi}}}.
\end{eqnarray}
This is the pNGB DM candidate we desire.
Another linear combination of $\eta_S$ and $\eta_\Phi$ is a massless Nambu-Goldstone boson eaten by a gauge boson.

After the spontaneous breaking of the gauge symmetry, the $\SUtwoL \times \UoneY \times \UoneX$ gauge fields acquire the mass terms
\begin{equation}
\mathcal{L}_{\mathrm{mass}}
\supset \frac{g^2 v^2}{4} W^{-, \mu} W^+_{\mu} + \frac{v^2}{8} (g^2 W^{3, \mu}
W^3_{\mu} + g'^2 B^{\mu} B_{\mu} - 2 g g' B^{\mu} W^3_{\mu}) +
\frac{g_X^2}{2}  (v_S^2 + 4 v_{\Phi}^2) X^{\mu} X_{\mu}.
\end{equation}
Taking into account the $B^\mu$-$X^\mu$ kinetic mixing and the $B^\mu$-$W^{3,\mu}$ mass mixing, the physical neutral gauge fields $(A^\mu, Z^\mu, Z'^\mu)$ can be derived through linear combinations of $(B^\mu, W^{3,\mu}, X^\mu)$~\cite{Babu:1997st}.
The electromagnetic field $A^\mu$ is massless, while the masses for the $Z$ and $Z'$ bosons are given by~\cite{Chun:2010ve}
\begin{equation}
m_{Z}^{2}=\frac{v^{2}}{4}(g^{2}+g'^{2})(1+\hat{s}_\mathrm{W}t_{\varepsilon}t_{\xi}),\quad
m_{Z'}^{2}=\frac{g_{X}^{2}(v_{S}^{2}+4v_{\Phi}^{2})}{c_{\varepsilon}^{2}(1+\hat{s}_\mathrm{W}t_{\varepsilon}t_{\xi})},
\end{equation}
where $t_\varepsilon \equiv \tan \varepsilon$, $\hat{s}_\mathrm{W} \equiv \sin \hat{\theta}_\mathrm{W}$, $\hat{c}_\mathrm{W} \equiv \cos \hat{\theta}_\mathrm{W}$, and $\hat{\theta}_\mathrm{W} \equiv \tan^{-1} (g'/g)$.
We define $r \equiv m_{Z'}^2/m_Z^2$, and $t_\xi \equiv \tan \xi$ can be expressed as~\cite{Lao:2020inc}
\begin{equation}
t_{\xi}=\frac{2\hat{s}_\mathrm{W}t_{\varepsilon}}{1-r}\left[1+\sqrt{1-r\left(\frac{2\hat{s}_\mathrm{W}t_{\varepsilon}}{1-r}\right)^2}\,\right]^{-1}.
\end{equation}
Note that $Z'$ is an exotic neutral vector boson.

The interactions in this model have been explicitly discussed in Ref.~\cite{Liu:2022evb}.
Here, we briefly summarize the phenomenologically important couplings:
\begin{itemize}
\item The $h_i \chi\chi$ and $h_i h_j h_k$ couplings come from the scalar interactions.
\item The $h_i f f$ Yukawa couplings for any SM fermion $f$ arise from the SM Yukawa couplings and the mixings among the Higgs bosons.
\item The $h_i W W$, $h_i ZZ$, $h_i Z'Z'$, and  $h_i ZZ'$ couplings come from the covariant kinetic terms of the scalar fields.
\item  The $Z f f$, $Z' f f$, $Z \chi\chi$,  and $Z' \chi\chi$ neutral current couplings are induced by the gauge interactions and the kinetic and mass mixings of the gauge fields.
\item The $Z\chi h_i$ and $Z'\chi h_i$ couplings originating from the neutral current gauge interactions could lead to $\chi$ decays.
These couplings vanish in the $v_\Phi \to \infty$ limit, where $\chi$ is stable.
Thus, they must be greatly suppressed by a high UV scale $v_\Phi$ to give a sufficiently long lifetime for the DM candidate $\chi$.
\end{itemize}

\section{Phenomenological constraints}
\label{sec:phen}

Ten free parameters in this pNGB DM model can be chosen as $s_{\varepsilon}$, $\lambda_{HS}$, $\lambda_{H\Phi}$, $\lambda_{S\Phi}$, $v_{S}$, $v_{\Phi}$, $m_{\chi}$, $m_{h_{2}}$, $m_{h_{3}}$, and $m_{Z'}$.
As a UV completion of pNGB DM, we are interested in the parameter regions with $v \sim v_S \ll v_\Phi$, implying a mass hierarchy $m_{h_1} \sim m_{h_2} \ll m_{h_3} \sim m_{Z'}$.
According to effective field theory~\cite{Yu:2011by}, the tree-level of the spin-independent (SI) $\chi$-nucleon scattering cross section is obtained as~\cite{Liu:2022evb}
\begin{equation}
\sigma_{\chi N}^\mathrm{SI} \simeq \frac{m_N^4 m_\chi^4 [2 + 7(f_u^N + f_d^N + f_s^N)]^2 (\lambda _{H\Phi }\lambda _{S\Phi } - \lambda _\Phi \lambda _{HS} + 2\lambda _{HS}\lambda _{S\Phi } - 2\lambda _S \lambda _{H\Phi })^2}{1296\pi  v^4 v_\Phi^4 (m_N + m_\chi)^2 (\lambda _H\lambda _S\lambda _\Phi  + 2\lambda _{HS}\lambda _{H\Phi }\lambda _{S\Phi } - \lambda _S\lambda _{H\Phi }^2 - \lambda _\Phi \lambda _{HS}^2 - \lambda _H\lambda _{S\Phi }^2)^2},
\end{equation}
where $m_N$ represents the nucleon mass and $f_{u,d,s}^{N}$ are nucleon form factors~\cite{Ellis:2000ds}.
Note that $\sigma_{\chi N}^\mathrm{SI}$ is highly suppressed by $v_\Phi^{-4}$.
For $v_\Phi \lesssim 10^5~\si{GeV}$, $\sigma_{\chi N}^\mathrm{SI}$ is typically below $10^{-50}~\si{cm^2}$~\cite{Liu:2022evb}.
Therefore, direct detection experiments would hardly probe the pNGB DM candidate $\chi$.

As mentioned above, the $Z\chi h_i$ and $Z'\chi h_i$ couplings result in $\chi$ decays.
For $m_\chi \ll m_{h_3} \sim m_{Z'}$, the pNGB DM candidate $\chi$ could decay via $\chi \to h_i^{(*)} Z^{(*)}$ and $\chi \to h_i^{(*)} Z'^{*}$, where $h_3$ and $Z'$ are off shell, but $Z$, $h_1$, and $h_2$ can be either on or off shell.
These processes are all suppressed by the UV scale $v_\Phi$.
In order to explain dark matter in the present universe, $\chi$ must have a very long lifetime.
A conservative bound on the DM lifetime, \textit{i.e.}, $\tau_{\chi} \gtrsim 10^{27}~\si{s}$, has been given by Fermi-LAT $\gamma$-ray observations of nearby dwarf spheroidal galaxies~\cite{Baring:2015sza}.
This puts a strong constraint on $v_\Phi$.

In this model, $\chi \chi$ annihilation channels involve $h_i h_j$ ($i,j = 1,2$), $W^+W^-$, $ZZ$, and $f\bar{f}$.
Contrary to $\chi$-nucleon scattering, $\chi \chi$ annihilation has no particular suppression; thus, the observed DM relic abundance can be easily obtained via the freeze-out mechanism.
The $\chi \chi$ annihilation processes at the present would induce $\gamma$ rays, which can be probed by indirect detection experiments.

Moreover, the couplings of the SM-like Higgs boson $h_1$ to $W$, $Z$, and the SM fermions deviate from the SM. 
Exotic decays $h_1 \to \chi\chi$, $h_1 \to \chi Z$, and $h_1 \to h_2 h_2$ may occur for $m_{h_1} > 2m_\chi$, $m_{h_1} > m_\chi + m_Z$, and $m_{h_1} > 2m_{h_2}$, respectively.
Therefore, the LHC Higgs measurements have put some constraints on the parameter space of this model.
In addition, the exotic Higgs boson $h_2$ could be directly produced at high energy hadron colliders.
The $h_2$ production processes include gluon-gluon fusion $gg\to h_2$, vector boson fusion $WW/ZZ\to h_2$, and $h_2$ production associated with $W$, $Z$, $t\bar{t}$, or $b\bar{b}$, while the $h_2$ decay channels involve $t\bar{t}$, $b\bar{b}$, $W^+W^-$, $ZZ$, $\gamma\gamma$, $Z\gamma$, etc.
These processes lead to detectable signals of $h_2$ at the LHC and the Tevatron.
Nonetheless, the constraints from LHC and Tevatron direct searches can be evaded if the $h$ component in $h_2$ is small and/or $h_2$ is heavy.

In order to study the phenomenological constraints on the pNGB DM model, we use \texttt{FeynRules~2}~\cite{Alloul:2013bka} to implement the model and generate model files for the numerical package \texttt{micrOMEGAs~5.2}~\cite{Belanger:2020gnr}.
Utilizing \texttt{micrOMEGAs}, we calculate the $\chi$ lifetime $\tau_\chi$, the DM relic abundance $\Omega_\chi h^2$, and the $\chi\chi$ annihilation cross section $\langle {\sigma}_{\mathrm{ann}}v\rangle$.
\texttt{HiggsSignals~2}~\cite{Bechtle:2014ewa} is applied to test whether the properties of the $h_1$ boson are consistent with the LHC measurements, while \texttt{HiggsBounds~5}~\cite{Bechtle:2015pma} is employed to constrain the $h_2$ boson by LHC and Tevatron direct searches.

A random parameter scan is carried out within the following ranges of the free parameters:
\begin{eqnarray}
&& 10^{-3} < |s_\varepsilon| <0.9,\qquad 10^{-2}<|\lambda_{HS}|,\ |\lambda_{H\Phi}|,\ |\lambda_{S\Phi}|<1,
\nonumber\\
&& 10^{8}\ \mathrm{GeV}<v_{\Phi}<10^{15}\ \mathrm{GeV},\qquad 10^{7}\ \mathrm{GeV}<m_{h_{3}},\ m_{Z'}<10^{16}\ \mathrm{GeV},\nonumber\\
&& 10\ \mathrm{GeV}< v_{S},\ m_{h_2},\ m_\chi <10^{4}\ \mathrm{GeV}.
\end{eqnarray}
The parameter points are randomly generated by uniform distributions on the logarithmic scale.
We further demand the induced couplings $g_X$, $\lambda_H$,  $\lambda_S$,  and $\lambda_\Phi$ lying from $10^{-2}$ to $1$.
Because we have required $v_\Phi > 10^8~\si{GeV}$, the direct detection constraints~\cite{PandaX-4T:2021bab, Aalbers:2022fxq, XENON:2023sxq} are totally irrelevant.

The following criteria are used to select the parameter points.
\begin{enumerate}
\item In order to guarantee the vacuum stability, the following conditions for the scalar potential from copositivity criteria~\cite{Kannike:2012pe} should be satisfied:
\begin{equation}
\lambda_H \geq 0,\quad
\lambda_S \geq 0,\quad
\lambda_{\Phi} \geq  0,
\end{equation}
\begin{equation}
a_{HS} \equiv \lambda_{H S} + \sqrt{\lambda_H \lambda_S} \geq 0,\quad
a_{H\Phi} \equiv \lambda_{H \Phi} + \sqrt{\lambda_H
  \lambda_{\Phi}} \geq 0,\quad
a_{S\Phi} \equiv \lambda_{S \Phi} +
  \sqrt{\lambda_S \lambda_{\Phi}} \geq 0,
\end{equation}
\begin{equation}
\sqrt{\lambda_H \lambda_S \lambda_{\Phi}} + \lambda_{H S} \sqrt{\lambda_{\Phi}} + \lambda_{H \Phi} \sqrt{\lambda_S} + \lambda_{S \Phi}
  \sqrt{\lambda_H} + \sqrt{2 a_{HS} a_{H\Phi} a_{S\Phi}} \geq  0.
\end{equation}
\item The lifetime of the pNGB DM particle $\chi$ should satisfy the Fermi-LAT bound $\tau_{\chi} \gtrsim 10^{27}~\si{s}$~\cite{Baring:2015sza}.
\item The DM relic abundance $\Omega_\chi h^2$ should be in the $3\sigma$ range of the Planck value $\Omega_\mathrm{DM} h^2 = 0.1200\pm 0.0012$~\cite{Planck:2018vyg}.
\item The total $\chi\chi$ annihilation cross section $\langle {\sigma}_{\mathrm{ann}}v\rangle$ should not be excluded by the upper limits at the 95\% confidence level (C.L.) given by the combined Fermi-LAT and MAGIC $\gamma$-ray observations of dwarf spheroidal galaxies in the $b\bar{b}$ channel~\cite{MAGIC:2016xys}.
\item The signal strengths of the SM-like Higgs boson $h_1$ should be consistent with the LHC Higgs measurements at 95\% C.L. based on the \texttt{HiggsSignals} calculation.
\item The exotic Higgs boson $h_2$ should not be excluded at 95\% C.L. by the direct searches at the LHC and the Tevatron according to \texttt{HiggsBounds}.
\end{enumerate}

\begin{figure}[!t]
\centering
\subfigure[$m_{\chi}$-$\langle {\sigma}_{\mathrm{ann}}v\rangle$ plane\label{fig:mchi_sv}]{\includegraphics[width=0.485\textwidth]{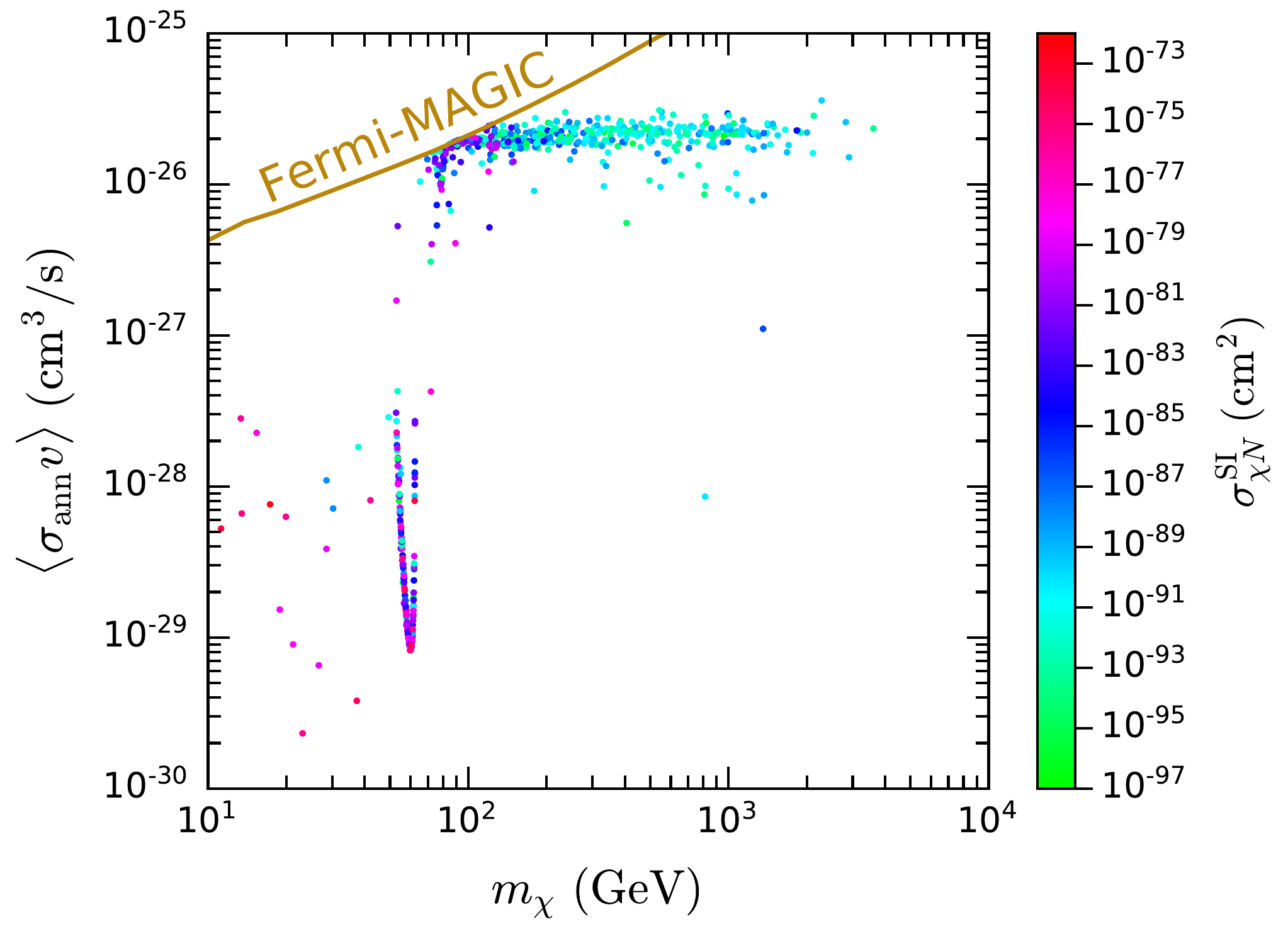}}
\hspace{.01\textwidth}
\subfigure[$v_{\Phi}$-$\lambda_{\phi}$ plane\label{fig:vPhi_lamPhi}]{\includegraphics[width=0.47\textwidth]{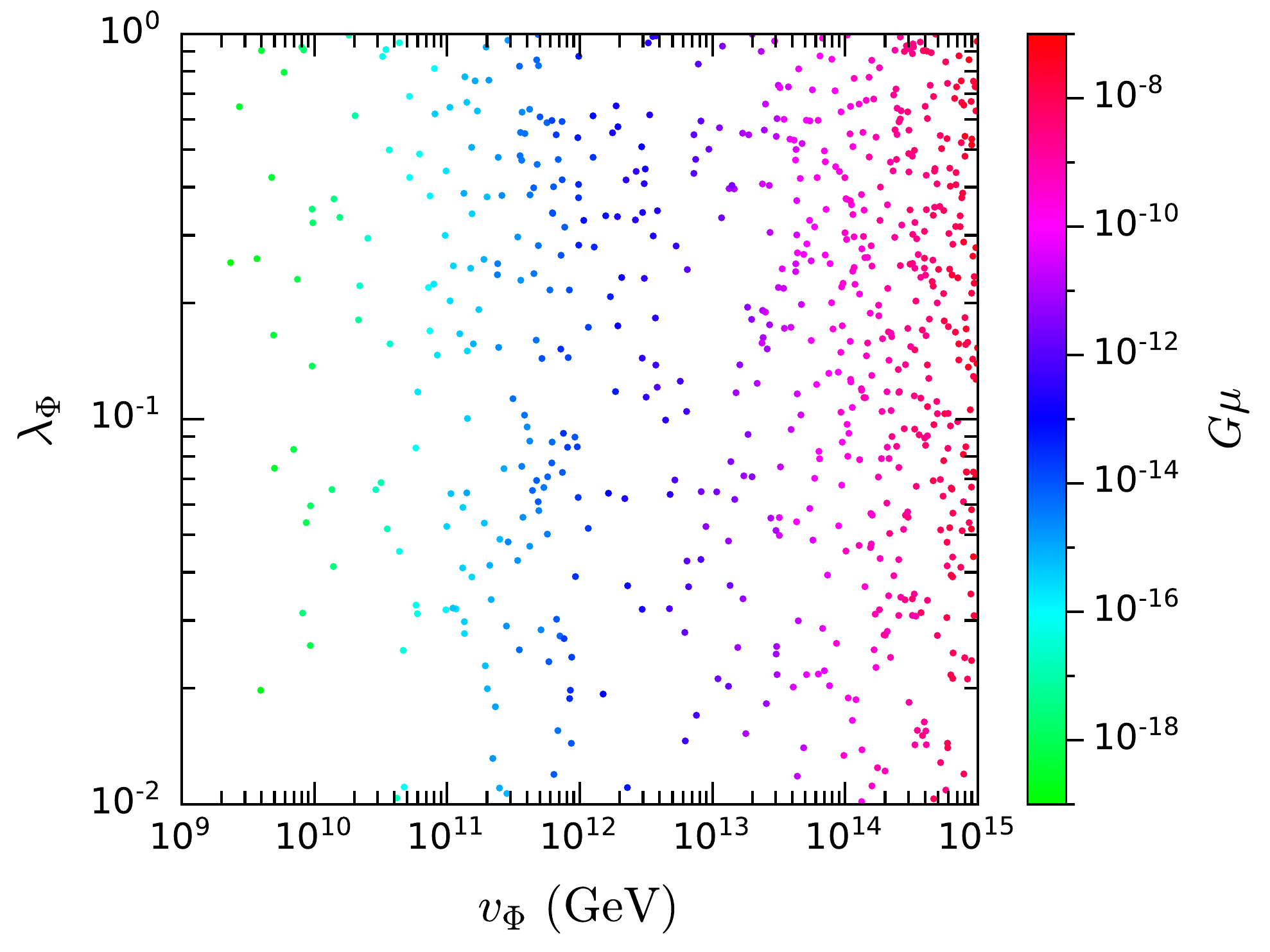}}
\caption{Selected parameter points projected onto the $m_{\chi}$-$\langle {\sigma}_{\mathrm{ann}}v\rangle$ (a) and $v_{\Phi}$-$\lambda_{\phi}$ (b) planes, with colors denoting $\sigma_{\chi N}^\mathrm{SI}$ and $G\mu$, respectively.\label{fig:scan}}
\end{figure}

In Fig.~\ref{fig:scan}, the selected parameter points are projected onto the $m_{\chi}$-$\langle {\sigma}_{\mathrm{ann}}v\rangle$ and $v_\Phi$-$\lambda_{\Phi}$ planes.
In order to achieve the correct DM relic abundance, most of the parameter points have a canonical annihilation cross section $\langle {\sigma}_{\mathrm{ann}}v\rangle \simeq \num{2e-26}~\si{cm^3/s}$, as shown in Fig.~\ref{fig:mchi_sv}.
Deviations from this value are mainly due to the $h_1$ and $h_2$ resonance effects.
Such resonances could greatly increase the $\chi\chi$ annihilation cross section at the freeze-out epoch,
but the present $\langle {\sigma}_{\mathrm{ann}}v\rangle$ would be rather small, because the center-of-mass energy of $\chi\chi$ at low velocities today would not be close to the resonance centers.
Obviously, the deep valley around $m_{h_1}/2 \simeq 63~\si{GeV}$ in Fig.~\ref{fig:mchi_sv} corresponds to the $h_1$ resonance.
The $h_2$ mass is not fixed, and the $h_2$ resonance effect is probably responsible for other parameter points with $\langle {\sigma}_{\mathrm{ann}}v\rangle$ that deviate from the canonical value.
From Fig.~\ref{fig:mchi_sv}, we find that in order to cover all the parameter points, the sensitivity of indirect detection experiments needs to be improved by three orders of magnitude.
The color axis in Fig.~\ref{fig:mchi_sv} indicates the tree-level SI DM-nucleon scattering cross section $\sigma_{\chi N}^\mathrm{SI}$, whose values are far too low to be reached by direct detection experiments.

Fig.~\ref{fig:vPhi_lamPhi} shows that the UV scale $v_\Phi$ should be higher than $10^{9}~\si{GeV}$\footnote{Compared with the scan in Ref.~\cite{Liu:2022evb} that found $v_\Phi > 10^{10}~\si{GeV}$, we perform more searches in the parameter space and find some parameter points with $10^{9}~\si{GeV} < v_\Phi < 10^{10}~\si{GeV}$.}, because of the bound on the DM lifetime.
Since a smaller $v_{\Phi}$ tends to give a shorter lifetime of $\chi$, the parameter points with smaller $v_{\Phi}$ have higher probabilities to be rejected by the DM lifetime bound than those with larger $v_{\Phi}$.
Considering the DM detection experiments and the measurements and searches at colliders, GW experiments provide a complementary way to test this model.
In the following sections, we will discuss the potential GW signals associated with the selected parameter points.

\section{Gravitational waves generated by cosmic strings}
\label{sec:CS}

The spontaneous breaking of the $\UoneX$ gauge symmetry at the high UV scale $v_\Phi$ is expected to induce cosmic strings, which are one-dimensional topological defects concentrated with energies of the scalar and gauge fields~\cite{Hindmarsh:1994re}.
The key quantity of a cosmic string is its tension $\mu$, which is the energy per unit length.
According to the analysis based on the Abelian Higgs model~\cite{Hill:1987qx}, the tension in the pNGB DM model can be estimated as follows:
\begin{eqnarray}\label{eq:Gmu}
\mu \simeq \begin{cases}
1.19 \pi v_{\Phi}^2 b^{-0.195}, & 0.01<b<100,\\[.4em]
\dfrac{2.4\pi v_{\Phi}^2}{\mathrm{ln}\ b} ,  & b>100,
\end{cases}
\end{eqnarray}
with $b \equiv 8 g_X^2 / \lambda_{\Phi}$.
Because $\mu \propto v_\Phi^2$, a high UV scale $v_\Phi$ suggested by the Fermi-LAT bound on the $\chi$ lifetime would lead to cosmic strings with high tension.
As we will see below, this would result in an SGWB with a high energy density $\propto G \mu^2$.

The dimensionless quantity $G\mu$, where $G$ is the Newtonian constant of gravitation, is commonly used to describe the tension of cosmic strings.
According to Eq.~\eqref{eq:Gmu}, we calculate $G\mu$ for the selected parameter points, and the result is shown on the $v_{\Phi}$-$\lambda_{\phi}$ plane in Fig.~\ref{fig:vPhi_lamPhi} with the color axis ranging from $10^{-19}$ to $10^{-7}$.
The positive correlation between $G\mu$ and $v_\Phi$ is clearly demonstrated.

A network of cosmic strings would be formed in the early universe after the spontaneous breaking of the $\UoneX$ gauge symmetry.
According to the analysis of string dynamics~\cite{Vilenkin:2000jqa}, the intersections of long strings could produce closed loops, whose size is smaller than the Hubble radius.
Cosmic string loops could further fragment into smaller loops or reconnect to long strings.
The relativistic oscillations of the loops due to their tension emit GWs, and the loops would shrink because of energy loss.
Moreover, the loops typically have special features called cusps and kinks, which could produce GW bursts~\cite{Damour:2000wa}.
Consequently, the energy of the cosmic string network is converted into the energy of GWs, and an SGWB is formed due to the incoherent superposition of GWs.

At the emission time $t_\mathrm{e}$, a cosmic string loop of length $L$ emits GWs with frequencies 
\begin{equation}\label{eq:f}
f_\mathrm{e} = \frac{2n}{L},
\end{equation}
where the integer $n=1,2,3,\cdots$ indicates the harmonic modes of the loop oscillation.
Denoting $P_n$ as the power of gravitational radiation for a mode $n$ in units of $G\mu^2$, the total power is given by~\cite{Vachaspati:1984gt}
\begin{equation}
P = G\mu^2\sum_n P_n.
\end{equation}
According to the numerical simulation of smoothed cosmic string loops, the averaged power spectra $P_n$ for loops in the radiation- and matter-dominated eras are obtained in Ref.~\cite{Blanco-Pillado:2017oxo}, including the contributions from cusps.
The total power can be characterized by the dimensionless quantity
\begin{equation}
\Gamma \equiv \frac{P}{G\mu^2} = \sum_n P_{n} ,
\end{equation}
which is estimated to be $\sim 50$ in the two eras~\cite{Blanco-Pillado:2017oxo}.
We will adopt this result for $P_{n}$ and $\Gamma = 50$ in the calculation of the GW spectrum below.

Defining $\mathsf{n}(L,t) \,\mathrm{d}L$ as the number density per physical volume of cosmic string loops with length $L$ at cosmic time $t$ in length interval $\mathrm{d} L$, the GW energy density $\rho_\mathrm{GW}$ induced by the cosmic string network per unit time at the emission time $t_\mathrm{e}$ can be expressed as
\begin{equation}
\frac{\mathrm{d}\rho_{\mathrm{GW}}}{\mathrm{d}t_\mathrm{e}}=G\mu^2\sum_n \int P_n\,\mathsf{n}(L,t_\mathrm{e})\,\mathrm{d}L.
\end{equation}
Utilizing Eq.~\eqref{eq:f}, we derive 
\begin{equation}
\frac{\mathrm{d}\rho_{\mathrm{GW}}}{\mathrm{d}t_\mathrm{e}\,\mathrm{d}f_\mathrm{e}}=G\mu^2\sum_n \frac{2n P_n}{f_\mathrm{e}^2}\,\mathsf{n}\!\left(\frac{2n}{f_\mathrm{e}},t_\mathrm{e}\right).
\end{equation}
Because of the cosmological redshift effect, the GW frequency $f$ observed at the present time $t_0$ differs from the emission frequency $f_\mathrm{e}$.
They are related by
\begin{equation}
f = \frac{a(t_\mathrm{e})}{a(t_0)}\,f_\mathrm{e} = \frac{f_\mathrm{e}}{1+z},
\end{equation}
where $a(t)$ is the cosmological scale factor and $z$ represents the redshift.
Setting $a(t_0)=1$, the GW energy at the present is given by $\rho_{\mathrm{GW}}(t_0) = \rho_{\mathrm{GW}}(t_\mathrm{e}) a^4(t_\mathrm{e})$, and we obtain
\begin{equation}\label{eq:drho/df}
\frac{\mathrm{d}\rho_{\mathrm{GW}}}{\mathrm{d}f}=G\mu^2\int_{t_*}^{t_0} a^5(t_\mathrm{e}) \sum_{n}\frac{2n P_n}{f^2}\,\mathsf{n}\!\left(\frac{2na(t_\mathrm{e})}{f},t_\mathrm{e}\right)\,\mathrm{d}t_\mathrm{e},
\end{equation}
where $t_*$ represents the time when the GW emissions start.
According to $\mathrm{d}t_\mathrm{e} = - [H(z)(1+z)]^{-1} \mathrm{d}z$, where $H(z)$ represents the Hubble expansion rate, we can rephrase Eq.~\eqref{eq:drho/df} as
\begin{equation}\label{eq:drho/df2}
\frac{\mathrm{d}\rho_{\mathrm{GW}}}{\mathrm{d}f}=G\mu^2\int_{0}^{z_*} \frac{1}{H(z)(1+z)^6} \sum_{n}\frac{2n P_n}{f^2}\,\mathsf{n}\!\left(\frac{2n}{f(1+z)},t(z)\right)\,\mathrm{d}z.
\end{equation}

The dimensionless quantity commonly used to characterize the SGWB is the spectrum of the GW energy density per logarithmic frequency interval divided by the critical density $\rho_\mathrm{c} = 3H_0^2/(8\pi G)$, \textit{i.e.},
\begin{equation}
	\Omega_{\mathrm{GW}}(f)=\frac{1}{\rho_\mathrm{c}}\frac{\mathrm{d}\mathrm{\rho_{\mathrm{GW}}}}{\mathrm{d}\ln f}=\frac{f}{\rho_\mathrm{c}}\frac{\mathrm{d}\mathrm{\rho_{\mathrm{GW}}}}{\mathrm{d}f}.
\end{equation}
The Hubble constant is usually expressed as $H_0 = 100h\ \mathrm{km}~ \mathrm{s}^{-1}~ \mathrm{Mpc}^{-1}$ with $h=0.674 \pm 0.005$~\cite{Planck:2018vyg}.
In order to avoid the uncertainty on the Hubble constant, one prefers to use $\Omega_{\mathrm{GW}}(f)h^2$.
For calculating Eq.~\eqref{eq:drho/df2}, we need to know $H(z)$.
In a flat $\Lambda\mathrm{CDM}$ universe, the Hubble rate is given by~\cite{Binetruy:2012ze}
\begin{equation}
H(z)=H_0\sqrt{\Omega_\mathrm{m}(1+z)^3 +  \Omega_\mathrm{r}(1+z)^4 \mathcal{G}(z) + \Omega_{\Lambda}},
\end{equation}
where $\Omega_\mathrm{m} = 0.315 \pm 0.007$~\cite{Planck:2018vyg}, $\Omega_\mathrm{r} = 1.68 \times (5.38\pm 0.15) \times 10^{-5}$~\cite{ParticleDataGroup:2022pth}, and $\Omega_{\Lambda} = 1 - \Omega_\mathrm{r} - \Omega_\mathrm{m}$ represent the energy densities of matter, radiation, and dark energy relative to the critical density at the present, respectively.
The function
\begin{equation}
\mathcal{G}(z)=\frac{g_{*}(z)}{g_{*}(0)} \left[\frac{g_{*s}(0)}{g_{*s}(z)}\right]^{4/3}
\end{equation}
takes account of the change in the number of radiation degrees of freedom between redshift $z$ and the present, where $g_{*}$ and $g_{*s}$ represent the effective numbers of relativistic degrees of freedom for the energy and entropy densities. 
Considering electron-positron annihilation at $z \simeq 10^9$ and the QCD phase transition at $z \simeq 2\times 10^{12}$, we have~\cite{Binetruy:2012ze}
\begin{equation}
\mathcal{G}(z) = \begin{cases}
	1, & z<10^9,\\
	0.83,  & 10^9<z<2\times 10^{12},\\
	0.39,  & z>2\times 10^{12}.
\end{cases}
\end{equation}

The last gradient that we need to compute the energy density spectrum of the SGWB is the loop number density distribution $\mathsf{n}(L,t)$.
There are various approaches for modeling $\mathsf{n}(L,t)$, which can lead to significant differences in the GW spectrum.
Here, we discuss two typical models for loop production.

The first model introduced by Blanco-Pillado, Olum, and Shlaer (BOS)~\cite{Blanco-Pillado:2013qja} makes use of the scaling nature of the cosmic string network and takes the horizon distance to be the only kinematic scale.
By extrapolating the loop production function found in numerical simulations of Nambu-Goto strings~\cite{Blanco-Pillado:2011egf}, $\mathsf{n}(L,t)$ is derived for any given cosmic time. 
Then, the loop number density produced in the radiation era can be approximated as
\begin{equation}
\mathsf{n}_\mathrm{r}(L,t)\simeq\frac{0.18}{t^{3/2}(L+\Gamma G \mu t)^{5/2}} \,\theta (0.1t-L),
\end{equation}
where $\theta(x)$ is the Heaviside step function.
The cosmic string loops produced in the matter era give a contribution of
\begin{equation}
\mathsf{n}_\mathrm{m}(L,t)\simeq\frac{0.27-0.45(L/t)^{0.31}}{t^2(L+\Gamma G \mu t)^2}\,\theta (0.18t-L).
\end{equation}
Moreover, there would be cosmic string loops produced in the radiation era and still surviving in the matter era. Their number density is given by
\begin{equation}
\mathsf{n}_{\mathrm{r}\to\mathrm{m}}(L,t)\simeq\frac{0.18t_{\mathrm{eq}}^{1/2}}{t^2(L+\Gamma G \mu t)^{5/2}}\,\theta(0.09t_{\mathrm{eq}}-\Gamma G \mu t-L),
\end{equation}
where $t_{\mathrm{eq}}=51.1 \pm 0.8~\mathrm{kyr}$ \cite{ParticleDataGroup:2022pth} represents the cosmic time at the matter-radiation equality.
Therefore, we have $\mathsf{n}(L,t) = \mathsf{n}_\mathrm{r}(L,t)$ for $t < t_\mathrm{eq}$ and $\mathsf{n}(L,t) = \mathsf{n}_\mathrm{m}(L,t) + \mathsf{n}_{\mathrm{r}\to\mathrm{m}}(L,t)$ for $t > t_\mathrm{eq}$.

The second model given by Lorenz, Ringeval, and Sakellariadou (LRS)~\cite{Lorenz:2010sm, Ringeval:2017eww} follows the Polchinski-Rocha approach~\cite{Polchinski:2006ee} and assumes that the number density distribution of produced loops per unit time $\mathcal{P}(L,t)$ is given by a power law in the form of
\begin{equation}
t^5\mathcal{P}(L,t) \propto \gamma^{2\chi-3},
\end{equation}
where $\gamma \equiv L/t$ is a dimensionless variable and $\chi$ is a parameter characterizing the power law.
Furthermore, they took into account the gravitational backreaction effect, which prevents loop production below a certain scale, by considering a different power law for $\gamma < \gamma_\mathrm{c}$ with $\chi$ changed to another parameter $\chi_\mathrm{c}$.
Here, $\gamma_\mathrm{c} \simeq 20 (G\mu)^{1+2\chi}$~\cite{Polchinski:2007rg} characterizes the length scale of gravitational backreaction.
The resulting loop number density distribution in scaling is found to be insensitive to the value of $\chi_\mathrm{c}$ and can be expressed as $\mathsf{n}(L,t)=t^{-4}N(\gamma)$, with the function $N(\gamma)$ approximately given by~\cite{Lorenz:2010sm}
\begin{equation}
N(\gamma)\simeq \begin{cases}
\dfrac{C}{(\gamma+\gamma_\mathrm{d})^{3-2\chi}}, & \gamma_\mathrm{d} < \gamma,\\[1em]
\dfrac{(3\nu-2\chi-1)C}{2(1-\chi)\gamma_\mathrm{d} \gamma^{2(1-\chi)}},  & \gamma_\mathrm{c} < \gamma < \gamma_\mathrm{d},\\[1em]
\dfrac{(3\nu-2\chi-1)C}{2(1-\chi)\gamma_\mathrm{d} \gamma_\mathrm{c}^{2(1-\chi)}},  & \gamma < \gamma_\mathrm{c}.
\end{cases}
\end{equation}
Here, $\gamma_\mathrm{d} = -\mathrm{d}L/\mathrm{d}t \simeq \Gamma G\mu$ represents the shrinking rate of cosmic string loops.
$\nu$ is the exponent in the relation between the scale factor and the cosmic time, \textit{i.e.}, $a(t)\propto t^{\nu}$, with $\nu=1/2$ in the radiation era and $\nu=2/3$ in the matter era.
The parameters $C$ and $\chi$ can be expressed as
\begin{equation}
C=C_\circ(1-\nu)^{3-p},\quad
\chi=1 - \frac{p}{2},
\end{equation}
where $C_\circ$ and $p$ are extracted by the  distribution of scaling loops from Nambu-Goto simulations~\cite{Ringeval:2005kr}:
\begin{eqnarray}
&& C_\circ = 0.21_{-0.12}^{+0.13},\quad
p=1.60^{+0.21}_{-0.15}\quad \text{(radiation era)};\\
&& C_\circ = 0.09_{-0.03}^{+0.03},\quad
p=1.41^{+0.08}_{-0.07}\quad \text{(matter era)}.
\end{eqnarray}

\begin{figure}[!t]
\centering
\subfigure[Estimated GW spectra of four BPs\label{fig:Omegah2_LRS_BOS}]{\includegraphics[width=0.48\textwidth]{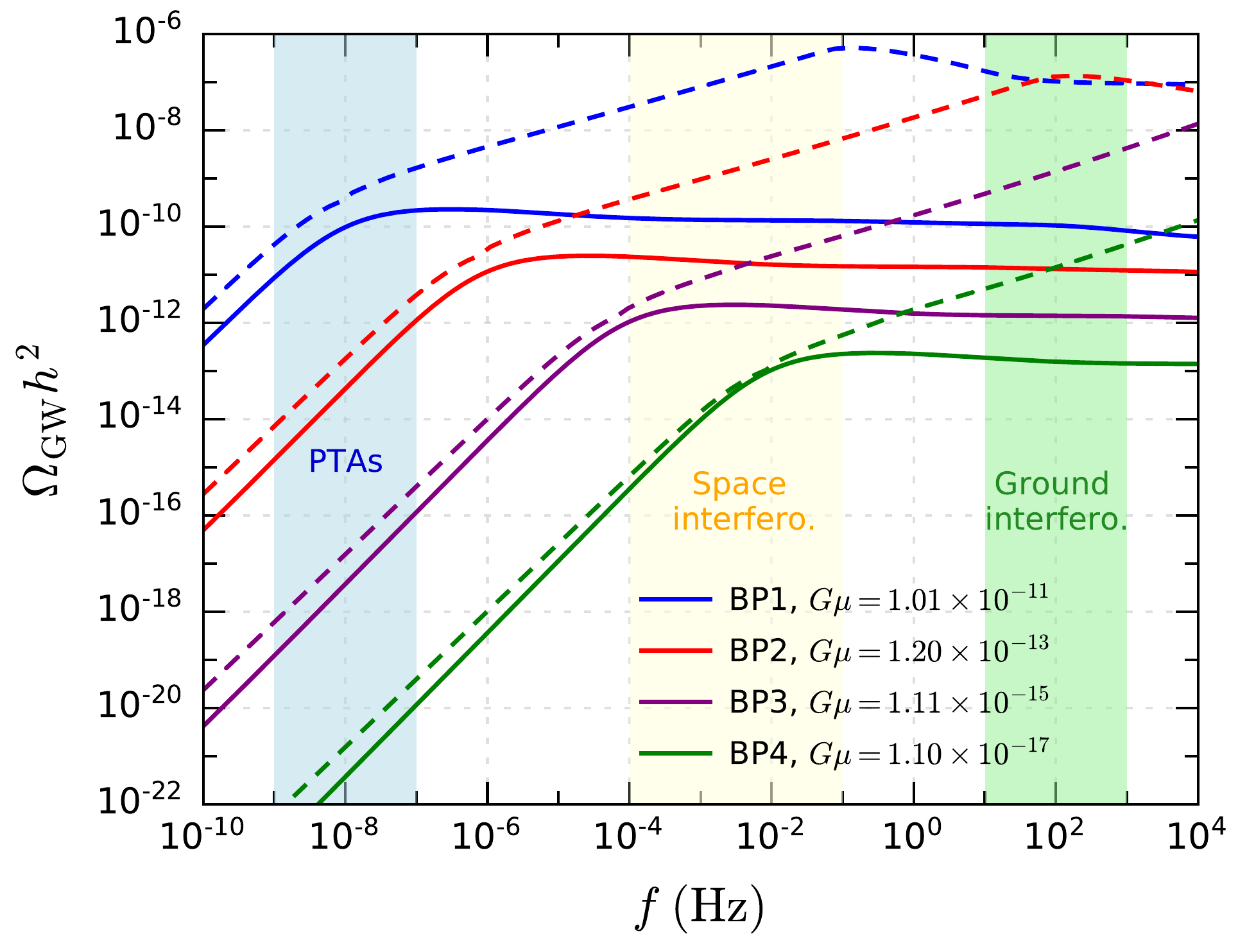}}
\hspace{.01\textwidth}
\subfigure[Sensitivity curves of GW experiments\label{fig:Omegah2_sens}]{\includegraphics[width=0.48\textwidth]{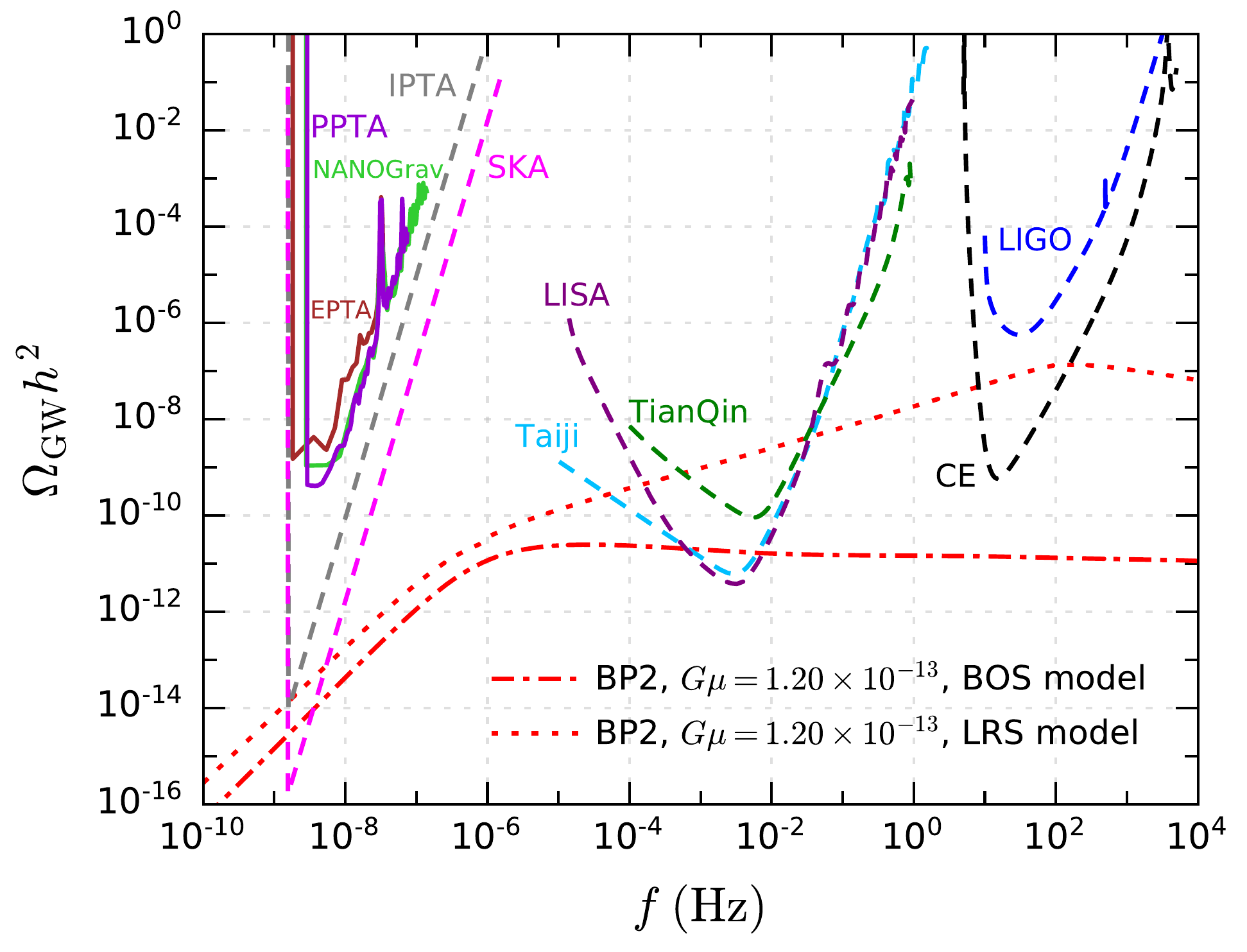}}
\caption{Estimated GW spectra of four BPs in the pNGB DM model and sensitivity curves of various experiments. In the left panel, the spectra of the SGWB estimated for the BOS (LRS) loop production model  are indicated by solid (dashed) lines. The sensitive frequency bands of PTAs, space interferometers, and ground interferometers are denoted by light blue, light yellow, and light green shaded bands, respectively. In the right panel, the upper limits from current GW experiments and the sensitivity curves of future GW experiments are indicated by solid and dashed lines, respectively, and the GW spectra of BP2 are shown for comparison.}
\end{figure}

Now, we can evaluate the energy density spectrum of the SGWB induced by the cosmic string network arising from the $\UoneX$ gauge symmetry breaking in the pNGB DM model.
From the selected parameter points, we choose four benchmark points (BPs), whose $G\mu$ varies from $\mathcal{O}(10^{-17})$ to $\mathcal{O}(10^{-11})$, to demonstrate the results.
Table~\ref{tab:BPs} presents the detailed information for the four BPs.
For each BP, we calculate the corresponding $\Omega_{\mathrm{GW}} h^2$ assuming the BOS and LRS loop production models, as plotted in Fig.~\ref{fig:Omegah2_LRS_BOS} with solid and dashed lines, respectively.

\begin{table}[!t]
\centering
\setlength\tabcolsep{.6em}
\renewcommand{\arraystretch}{1.3}
\caption{Information for four benchmark points.}\label{tab:BPs}
\begin{tabular}{ccccc}
\hline\hline
&  BP1   &  BP2  &  BP3  &  BP4\\
\hline
$v_S~(\mathrm{GeV})$     & 1953                & 2101                & 548.5                & 1388\\
$v_{\Phi}~(\mathrm{GeV})$&$1.335\times10^{13}$ &$1.939\times10^{12}$ &$1.969\times10^{11}$ &$3.179\times10^{10}$\\
$m_{\chi}~(\mathrm{GeV})$& 199.8               & 56.26               & 98.16               & 123.1\\
$m_{h_2}~(\mathrm{GeV})$ & 986.7               & 627.7               & 484.3               & 362.6\\
$m_{h_3}~(\mathrm{GeV})$ &$8.403\times10^{12}$ &$1.469\times10^{12}$ &$1.893\times10^{11}$ &$8.312\times10^{9}$\\
$m_{Z'}~(\mathrm{GeV})$  &$7.255\times10^{11}$ &$5.929\times10^{11}$ &$9.661\times10^{10}$ &$4.979\times10^{10}$\\
$\lambda_{H \Phi}$       &$-6.330\times10^{-2}$&$-3.786\times10^{-1}$&$-1.278\times10^{-2}$&$-6.114\times10^{-2}$\\
$\lambda_{S \Phi}$       &$-2.870\times10^{-1}$&$-5.416\times10^{-2}$&$2.813\times10^{-1}$ &$3.188\times10^{-2}$\\
$\lambda_{HS}$           &$3.259\times10^{-1}$ &$1.189\times10^{-1}$&$-1.750\times10^{-1}$ &$1.819\times10^{-2}$\\
$s_{\varepsilon}$        &$4.840\times10^{-3}$ &$3.222\times10^{-1}$ &$7.161\times10^{-2}$ &$1.929\times10^{-3}$\\
\hline
$\lambda_{\Phi}$         &$3.97 \times10^{-1}$ &$5.74\times10^{-1}$  &$9.24\times10^{-1}$  &$6.84\times10^{-2}$\\
$g_X$                    &$2.72 \times10^{-2}$ &$1.45\times10^{-1}$  &$2.45\times10^{-1}$  &$7.83\times10^{-1}$\\
$G\mu$                   &$1.01 \times10^{-11}$&$1.20\times10^{-13}$ &$1.11\times10^{-15}$ &$1.10\times10^{-17}$\\
${\Omega}_{\chi}h^2$     &$0.118$              &$0.121$              &$0.120$              &$0.119$            \\
${\sigma}_{\chi N}^{\mathrm{SI}}\ (\mathrm{cm}^2)$
&$1.38\times10^{-86}$ &$1.62\times10^{-86}$ &$1.59\times10^{-82}$ &$8.45\times10^{-77}$\\
$\langle {\sigma}_{\mathrm{ann}}v\rangle\ (\mathrm{cm^3/s})$
&$2.00\times10^{-26}$ &$2.87\times10^{-29}$ &$2.01\times10^{-26}$ &$1.71\times10^{-26}$\\
\hline
$\rho_{\mathrm{LISA}}$~(BOS)&$1.15\times10^{4}$&$1.48\times10^{3}$&$2.00\times10^{2}$&$3.97$\\
$\rho_{\mathrm{Taiji}}$~(BOS)&$7.26\times10^{3}$&$9.37\times10^{2}$&$1.26\times10^{2}$&$2.45$\\
$\rho_{\mathrm{TianQin}}$~(BOS)&$9.25\times10^{2}$ & $1.15\times10^{2}$ & $1.59\times10^{1}$ & $5.28\times10^{-1}$\\
$\rho_{\mathrm{CE}}$~(BOS)&$3.49\times10^{3}$&$4.33\times10^{2}$&$4.42\times10^{1}$&$5.48$\\
$\rho_{\mathrm{SKA}}$~(BOS)&$2.45\times10^{4}$&$4.61$&$3.93\times10^{-4}$&$3.90\times10^{-8}$\\
		\hline
$\rho_{\mathrm{LISA}}$~(LRS)&$1.15\times10^{7}$&$1.38\times10^{5}$&$1.28\times10^{3}$&$4.93$\\
$\rho_{\mathrm{Taiji}}$~(LRS)&$7.19\times10^{6}$&$8.57\times10^{4}$&$7.95\times10^{2}$&$3.05$\\
		$\rho_{\mathrm{TianQin}}$~(LRS)&$1.20\times10^{6}$ & $1.42\times10^{4}$ & $1.36\times10^{2}$ & $6.48\times10^{-1}$\\
$\rho_{\mathrm{CE}}$~(LRS)&$4.36\times10^{6}$&$2.18\times10^{6}$&$2.02\times10^{4}$&$2.11\times10^{2}$\\
$\rho_{\mathrm{SKA}}$~(LRS)&$1.13\times10^{5}$&$2.19\times10^{1}$&$1.87\times10^{-3}$&$1.86\times10^{-7}$\\
\hline\hline
\end{tabular}
\end{table}

In the BOS model, the high frequency behavior is controlled by the GW emissions during the radiation era~\cite{Blanco-Pillado:2013qja}.
As the integral contributed by $\mathsf{n}_\mathrm{r}(L,t)$ tends to be a constant at high frequencies, $\Omega_{\mathrm{GW}}h^2$ becomes quite flat in the high frequency regime, as indicated by the solid lines in Fig.~\ref{fig:Omegah2_LRS_BOS}.
For smaller $G\mu$, the total GW emission power is smaller, and cosmic string loops could survive longer, leading to more smaller loops radiating at higher frequencies.
Consequently, the GW spectrum moves downward and rightward as $G\mu$ decreases~\cite{Blanco-Pillado:2017oxo}.
Compared with the BOS model, the LRS model leads to GW spectra with far higher amplitudes, especially at high frequencies.
This is because the LRS model gives a very high number density of small loops in the $\gamma < \gamma_c$ regime, which significantly contribute to high frequency GWs~\cite{LIGOScientific:2017ikf, Auclair:2019wcv}.

\section{Constraints and sensitivity of GW experiments}
\label{sec:GW}

As shown in Fig.~\ref{fig:Omegah2_LRS_BOS}, the SGWB arising from cosmic strings extends over a very broad frequency range.
Thus, various GW experiments can probe it, and their sensitive bands are denoted by shaded regions in Fig.~\ref{fig:Omegah2_LRS_BOS}.
In this section, we investigate the constraints from current GW experiments on the pNGB DM model and the sensitivity in future experiments.

In Fig.~\ref{fig:Omegah2_sens}, we demonstrate the sensitivity curves of several GW experiments and compare them with the GW spectra of BP2 estimated by the BOS and LRS models.
For PTA experiments aiming at $10^{-9}\text{--} 10^{-7}~\mathrm{Hz}$, we show the 95\% C.L. upper limits from the European Pulsar Timing Array (EPTA) \cite{Lentati:2015qwp}, the North American Nanohertz Observatory for Gravitational Waves (NANOGrav)~\cite{NANOGRAV:2018hou}, and the Parkes Pulsar Timing Array (PPTA)~\cite{Shannon:2015ect}, as well as the projected strain noise spectra expressed in terms of the GW energy density~\cite{Schmitz:2020syl} (denoted as $\Omega_n h^2$ below) for the International Pulsar Timing Array (IPTA)~\cite{Hobbs:2009yy} and the Square Kilometer Array (SKA)~\cite{Janssen:2014dka}. 
For ground-based laser interferometers, whose sensitive band is $10^1\text{--} 10^{3}~\mathrm{Hz}$, we plot $\Omega_n h^2$ for the O5 runs of the Laser Interferometer Gravitational Wave Observatory (LIGO)~\cite{KAGRA:2013rdx} and for the Cosmic Explorer (CE)~\cite{LIGOScientific:2016wof}.
For future space-borne laser interferometers searching for GWs at $10^{-4}\text{--} 10^{-1}~\mathrm{Hz}$, we display $\Omega_n h^2$ for LISA~\cite{LISA:2017pwj}, TianQin~\cite{Liang:2021bde}, and Taiji~\cite{Ruan:2018tsw}.
For TianQin, $\Omega_n h^2$ is calculated from the detector noise power spectrum density and response function of the A/E channel given in Ref.~\cite{Liang:2021bde}.

\begin{figure}[!t]
\centering
\subfigure[BOS model\label{fig:vPhi_Gmu_BOS}]{\includegraphics[width=0.48\textwidth]{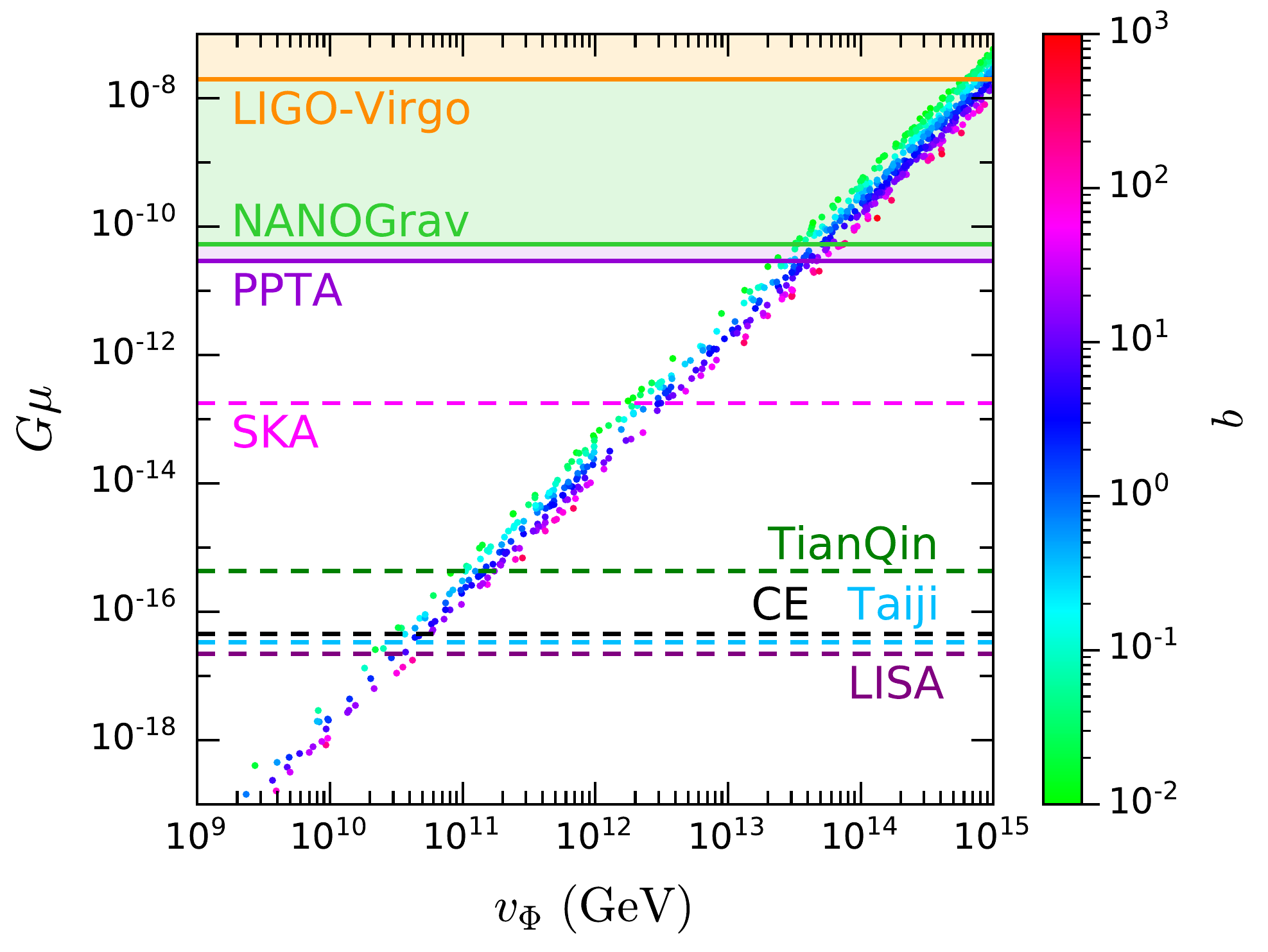}}
\hspace{.01\textwidth}
\subfigure[LRS model\label{fig:vPhi_Gmu_LRS}]{\includegraphics[width=0.48\textwidth]{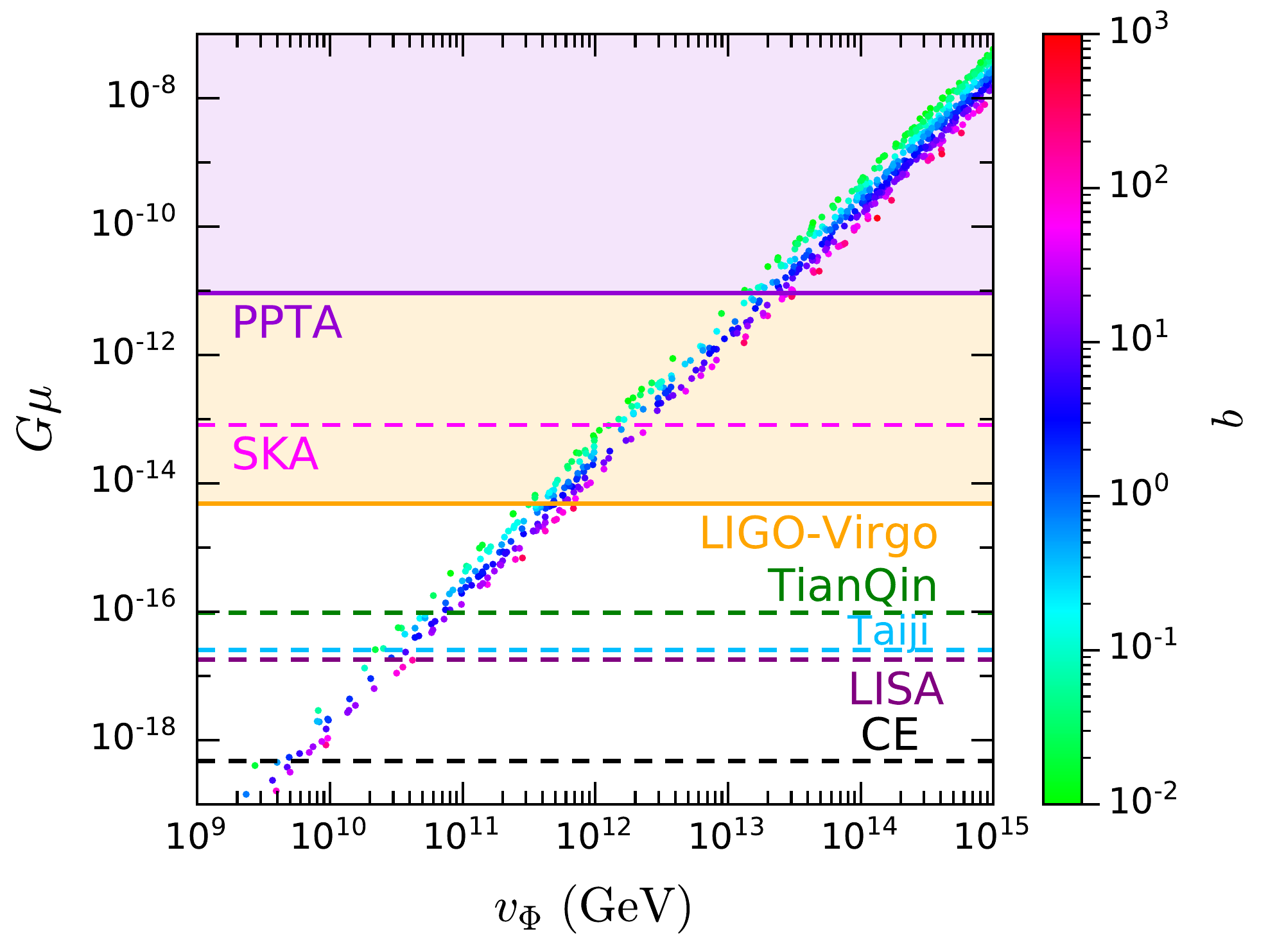}}
\caption{Selected parameter points projected onto the $v_{\Phi}$-$G\mu$ plane for the BOS~(a) and LRS~(b) models, with color axes corresponding to $b = 8 g_X^2 / \lambda_{\Phi}$. The constraints on $G\mu$ from current GW experiments are indicated by solid lines, and the expected upper limits in future experiments are denoted by dashed lines.\label{fig:vPhi_Gmu}}
\end{figure}

There are constraints on $G\mu$ from previous GW experiments, depending on the assumptions of the loop production models.
According to the LIGO-Virgo O3 data, assuming that the average numbers of cusps and kinks per loop oscillation are both $1$, the constraint for the BOS model at 95\% C.L. is $G\mu < 1.96 \times 10^{-8}$, while that for the LRS model is $G\mu < 4.83\times 10^{-15}$~\cite{LIGOScientific:2021nrg}. 
Compared with the BOS model, the LRS model predicts a higher amplitude in the frequency band of ground interferometers, leading to a far stronger constraint on  $G\mu$.
The 11-year NANOGrav data yield a 95\% C.L. upper limit of $G\mu < 5.3 \times 10^{-11}$ for the BOS model~\cite{NANOGRAV:2018hou}.
Utilizing the PPTA data over 15 years assuming a non-auto Hellings-Downs correlation~\cite{Bian:2022tju}, the 95\% C.L. constraints are estimated to be $G\mu < 2.88\times 10^{-11}$ for the BOS model\footnote{Another analysis of the PPTA data gives a 95\% C.L. upper bound of $G\mu < 5.1\times 10^{-10}$ for the BOS model~\cite{Chen:2022azo}.} and $G\mu < 9.12\times 10^{-12}$ for the LRS model.
These constraints and the selected parameter points are displayed on the $v_{\Phi}$-$G\mu$ plane for the BOS and LRS models in Figs.~\ref{fig:vPhi_Gmu_BOS} and \ref{fig:vPhi_Gmu_LRS}, respectively.
The color axes indicate the parameter $b = 8 g_X^2 / \lambda_{\Phi}$, which relates $v_\Phi$ to $G\mu$ through Eq.~\eqref{eq:Gmu}.
Because of the strong correlation between $G\mu$ and $v_\Phi$, the parameter points align around a straight line, and the dependence on $b$ scatters the points.
We find that the parameter points with $v_\Phi \gtrsim \num{5e13}~(\num{7e11})~\si{GeV}$ in the pNGB DM model have been excluded assuming the BOS (LRS) model for loop production.

Below, we evaluate the sensitivity of future GW experiments according to their expected $\Omega_{n}h^2$ shown in Fig.~\ref{fig:Omegah2_sens}.
For LISA, Taiji, CE, or SKA with an idealized auto-correlation measurement in an accessible frequency range $f_{\min} \leq f \leq f_{\max}$, the signal-to-noise ratio (SNR) $\rho$ can be estimated as~\cite{Thrane:2013oya, Schmitz:2020syl}
\begin{equation}\label{eq:SNR:1}
\rho = \sqrt{\mathcal{T}\int_{f_{\mathrm{min}}}^{f_{\mathrm{max}}}\left[\frac{\Omega_{\mathrm{GW}}(f)}{\Omega_{n}(f)}\right]^2 \mathrm{d}f},
\end{equation}
where $\mathcal{T}$ represents the practical observation time.
If the SNR $\rho$ turns out to be larger than an appropriate threshold, which will be set as $\rho_{\mathrm{thr}} = 10$ in the following, the predicted GW signal is likely to be detected.

For estimating the SNR of TianQin, we follow the strategy in Ref.~\cite{Liang:2021bde} with the null channel method, where the T channel is constructed to highly suppress the SGWB signal and the A and E channels are sensitive to the signal.
Assuming an ideal symmetric scenario where the A and E channels have the same $\Omega_{n}(f)$, the SNR can be evaluated as follows:
\begin{equation}\label{eq:SNR:2}
\rho_\mathrm{TianQin} = \sqrt{2\mathcal{T}\int_{f_{\mathrm{min}}}^{f_{\mathrm{max}}}\left[\frac{\Omega_{\mathrm{GW}}(f)}{\Omega_{n}(f)}\right]^2 \mathrm{d}f}.
\end{equation}
Because two channels are used, this expression has an additional factor $\sqrt{2}$ compared with Eq.~\eqref{eq:SNR:1}.

\begin{figure}[!t]
\centering
\subfigure[BOS model\label{fig:Gmu_SNR_BOS_space}]{\includegraphics[width=0.48\textwidth]{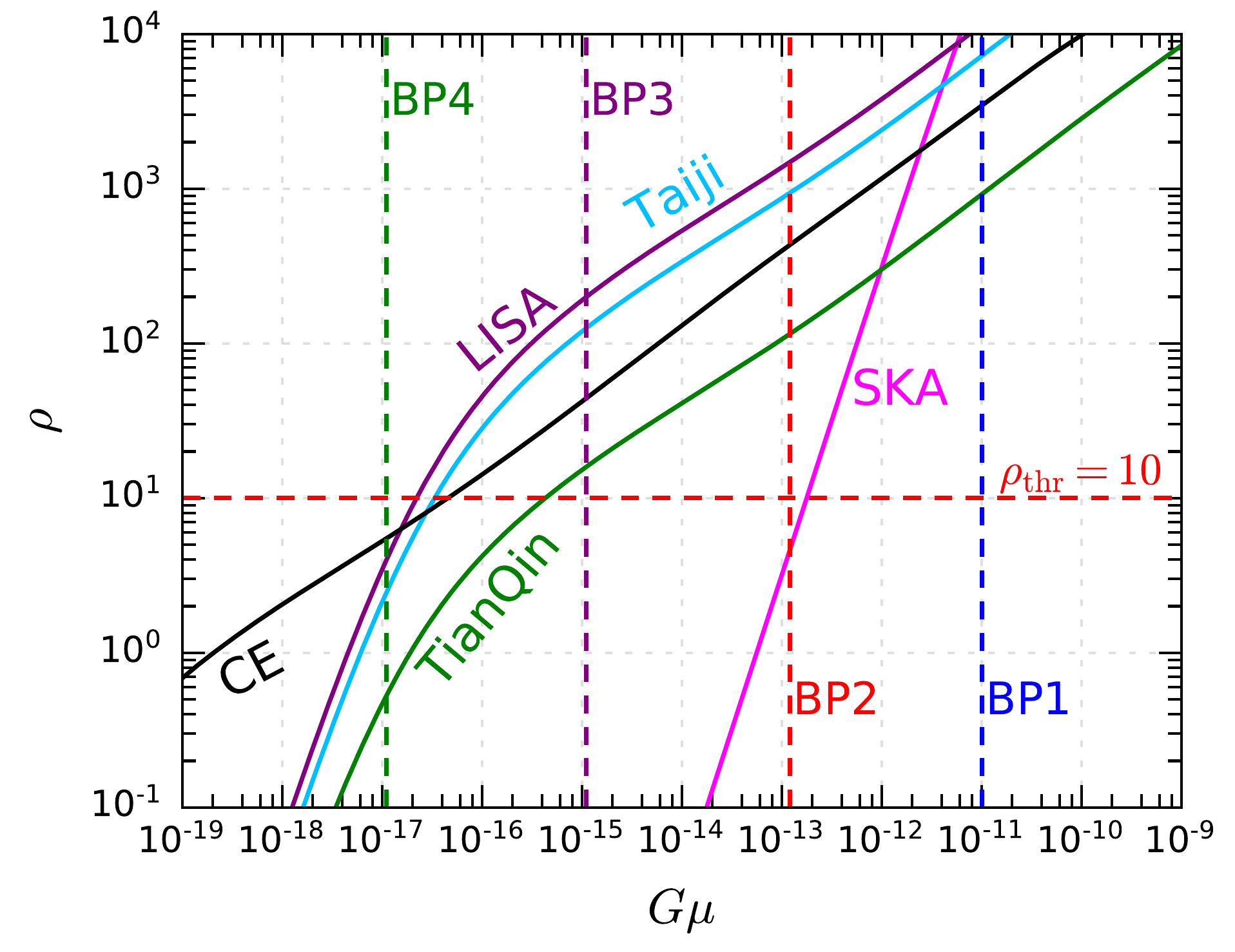}}
\hspace{.01\textwidth}
\subfigure[LRS model\label{fig:Gmu_SNR_LRS_space}]{\includegraphics[width=0.48\textwidth]{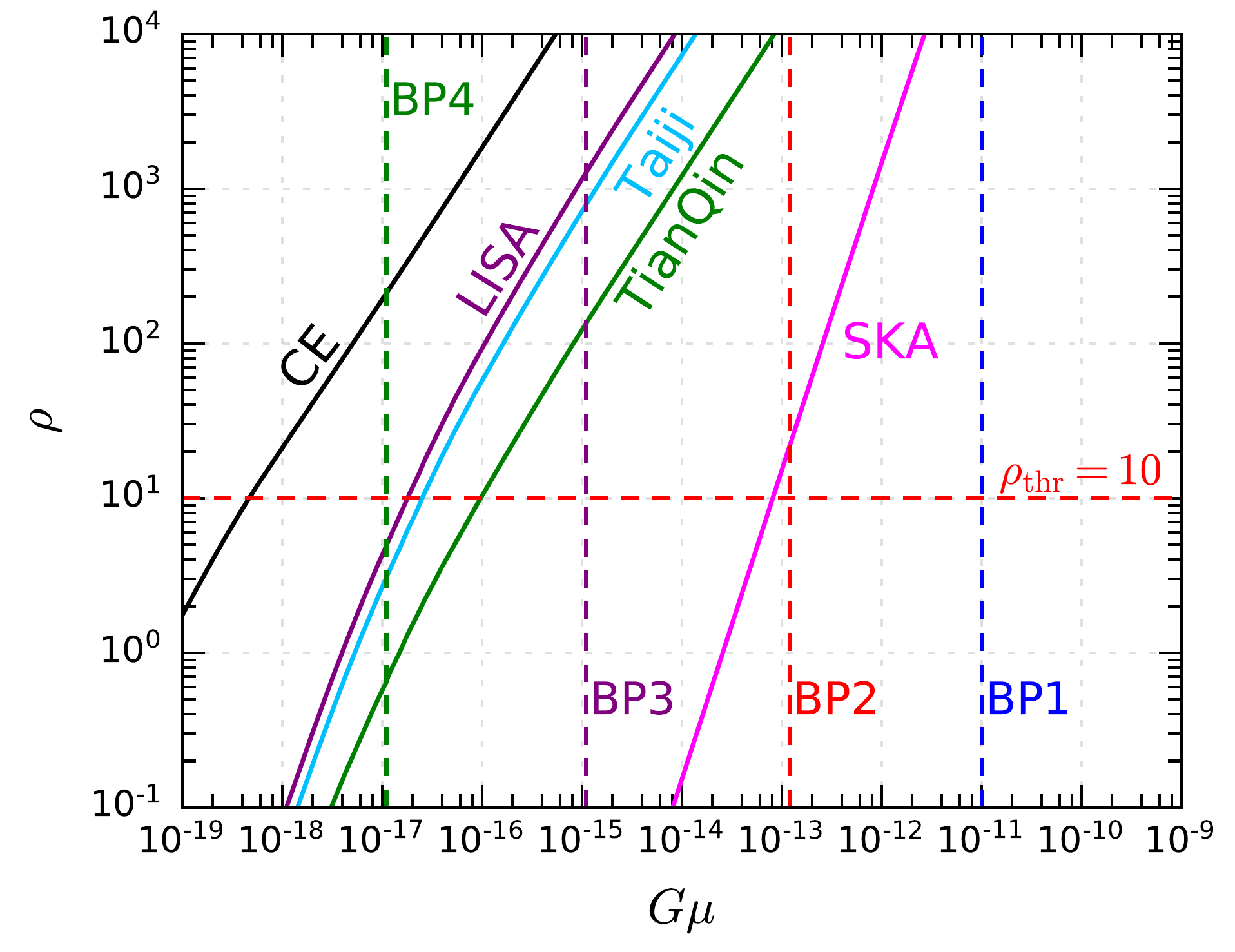}}
\caption{Estimated SNRs ($\rho$) with respect to $G\mu$ in future GW experiments LISA, Taiji, TianQin, CE, and SKA for the BOS (a) and LRS (b) models. The horizontal dashed lines correspond to the SNR threshold $\rho_\mathrm{thr} = 10$.
The vertical dashed lines indicate the values of $G\mu$ for the four BPs.}
\end{figure}

We assume that the observation time for LISA, Taiji, TianQin, or CE is $\mathcal{T} = 1~\si{yr}$, and that for SKA is  $\mathcal{T} = 10~\si{yr}$~\cite{Kuroyanagi:2012jf}.
Then, the SNR $\rho$ in these experiments is evaluated as a function of $G\mu$ for the BOS and LRS models, as shown in Figs.~\ref{fig:Gmu_SNR_BOS_space} and \ref{fig:Gmu_SNR_LRS_space}, respectively.
The estimated SNRs for the four BPs of the pNGB DM model are presented in Table~\ref{tab:BPs}.

\begin{table}[!t]
\setlength\tabcolsep{.6em}
\renewcommand{\arraystretch}{1.3}
\centering
\caption{Expected upper limits on $G\mu$ corresponding to $\rho_\mathrm{thr} = 10$.}\label{tab:constraints}
\begin{tabular}{cccccc}
\hline\hline
 & LISA   & Taiji  & TianQin & CE & SKA\\
\hline
BOS model & $2.21\times10^{-17}$ & $3.34\times10^{-17}$ & $4.28\times10^{-16}$ & $4.54\times10^{-17}$ & $1.77\times 10^{-13}$ \\
LRS model & $1.79\times10^{-17}$ 
 & $2.51\times10^{-17}$ & $9.67\times10^{-17}$ & $4.66\times10^{-19}$ & $8.09\times 10^{-14}$ \\
\hline\hline
\end{tabular}
\end{table}

\begin{figure}[!t]
\centering
\subfigure[$\rho_\mathrm{LISA}$ for the BOS model\label{fig:vPhi_SNR_BOS}]{\includegraphics[width=0.48\textwidth]{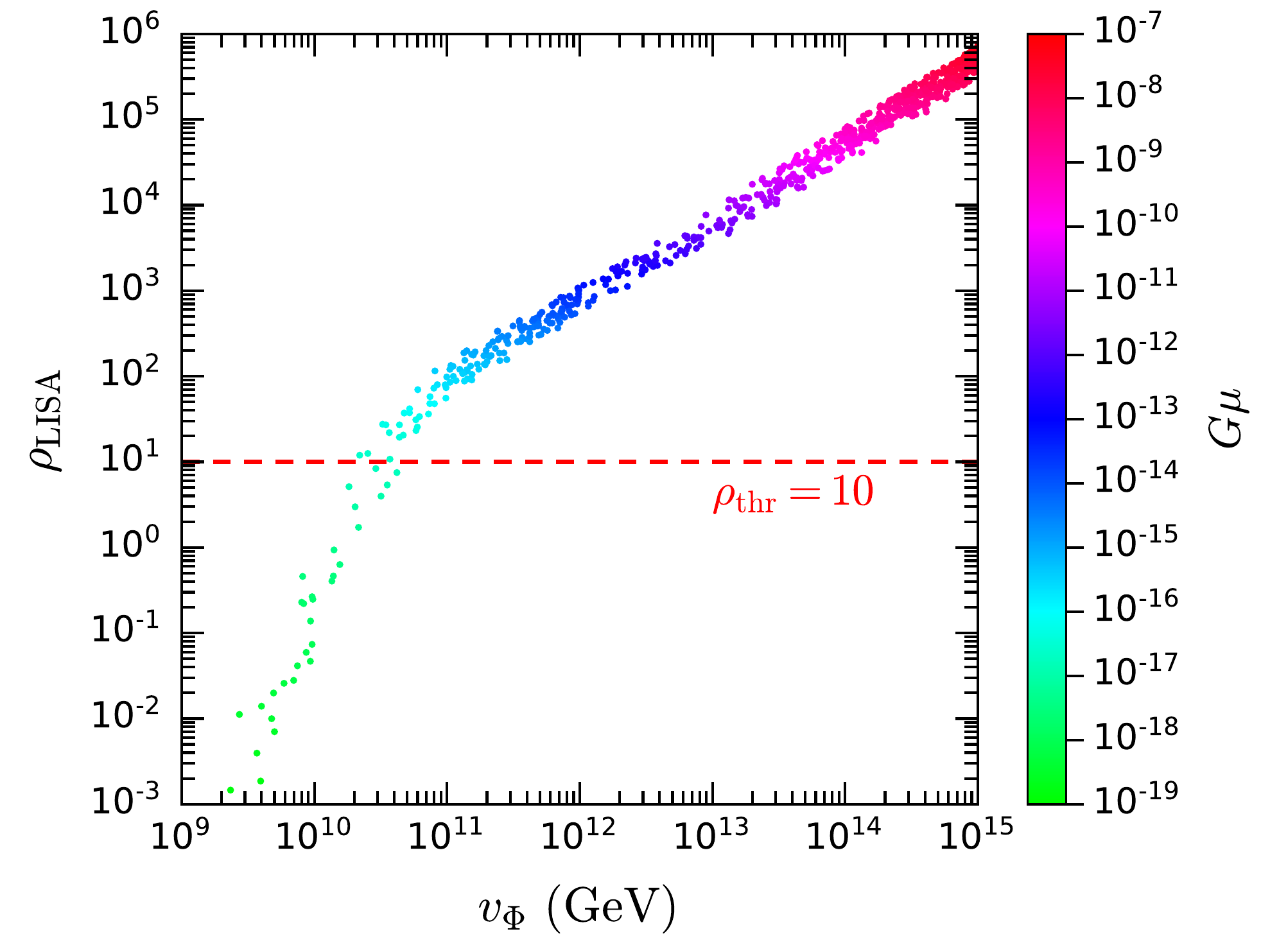}}
\hspace{.01\textwidth}
\subfigure[$\rho_\mathrm{CE}$ for the LRS model\label{fig:vPhi_SNR_LRS}]{\includegraphics[width=0.48\textwidth]{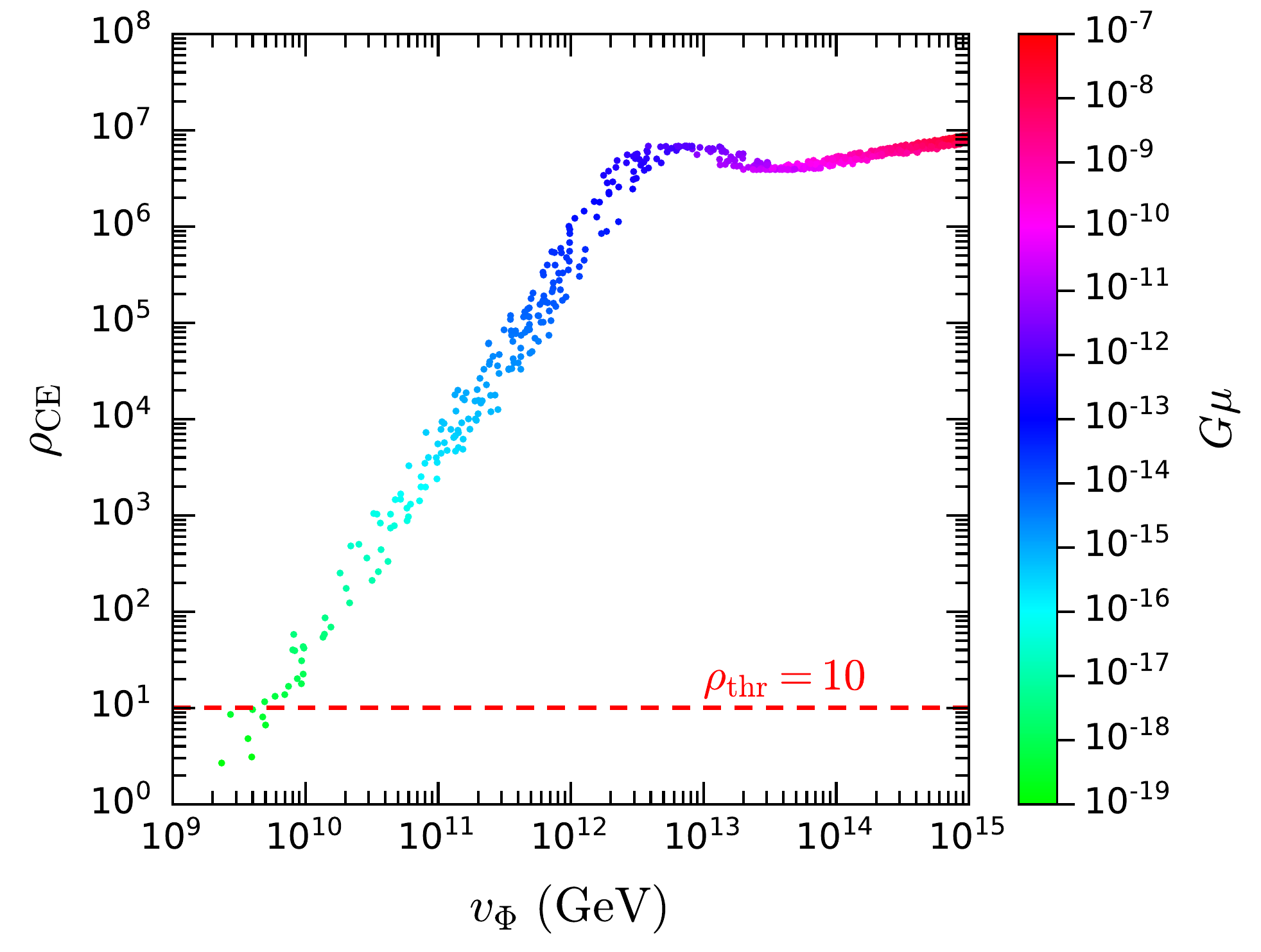}}
\caption{Selected parameter points projected onto the $v_{\Phi}$-$\rho_{\mathrm{LISA}}$ plane for the BOS model (a) and onto the $v_{\Phi}$-$\rho_{\mathrm{CE}}$ plane for the LRS model (b), with color axes corresponding to $G\mu$.
The horizontal dashed lines indicate the SNR threshold $\rho_\mathrm{thr} = 10$.\label{fig:vPhi_SNR}}
\end{figure}

Taking the SNR threshold to be $\rho_\mathrm{thr} = 10$, the expected upper limits on $G\mu$ for LISA, Taiji, TianQin, CE, and SKA are presented in Table~\ref{tab:constraints}.
These upper limits are plotted in Fig.~\ref{fig:vPhi_Gmu} for comparison with previous experimental results.
For the BOS model, the most sensitive experiments are LISA, Taiji, and CE, which have comparable sensitivities and could probe $v_\Phi$ down to $\sim\num{2e10}~\si{GeV}$.
For the LRS model, CE has the highest sensitivity, and the parameter points with $v_\Phi$ down to $\sim\num{5e9}~\si{GeV}$ could be detected.
In order to clearly demonstrate the sensitivity of future GW experiments to the pNGB DM model, we also project the selected parameter points onto the $v_\phi$-$\rho_\mathrm{LISA}$ plane for the BOS model and onto the $v_\phi$-$\rho_\mathrm{CE}$ plane for the LRS model in Fig.~\ref{fig:vPhi_SNR}, where $\rho_\mathrm{LISA}$ and $\rho_\mathrm{CE}$ represent the SNRs of LISA and CE, respectively.

\begin{figure}[!t]
\centering
\includegraphics[width=0.48\textwidth]{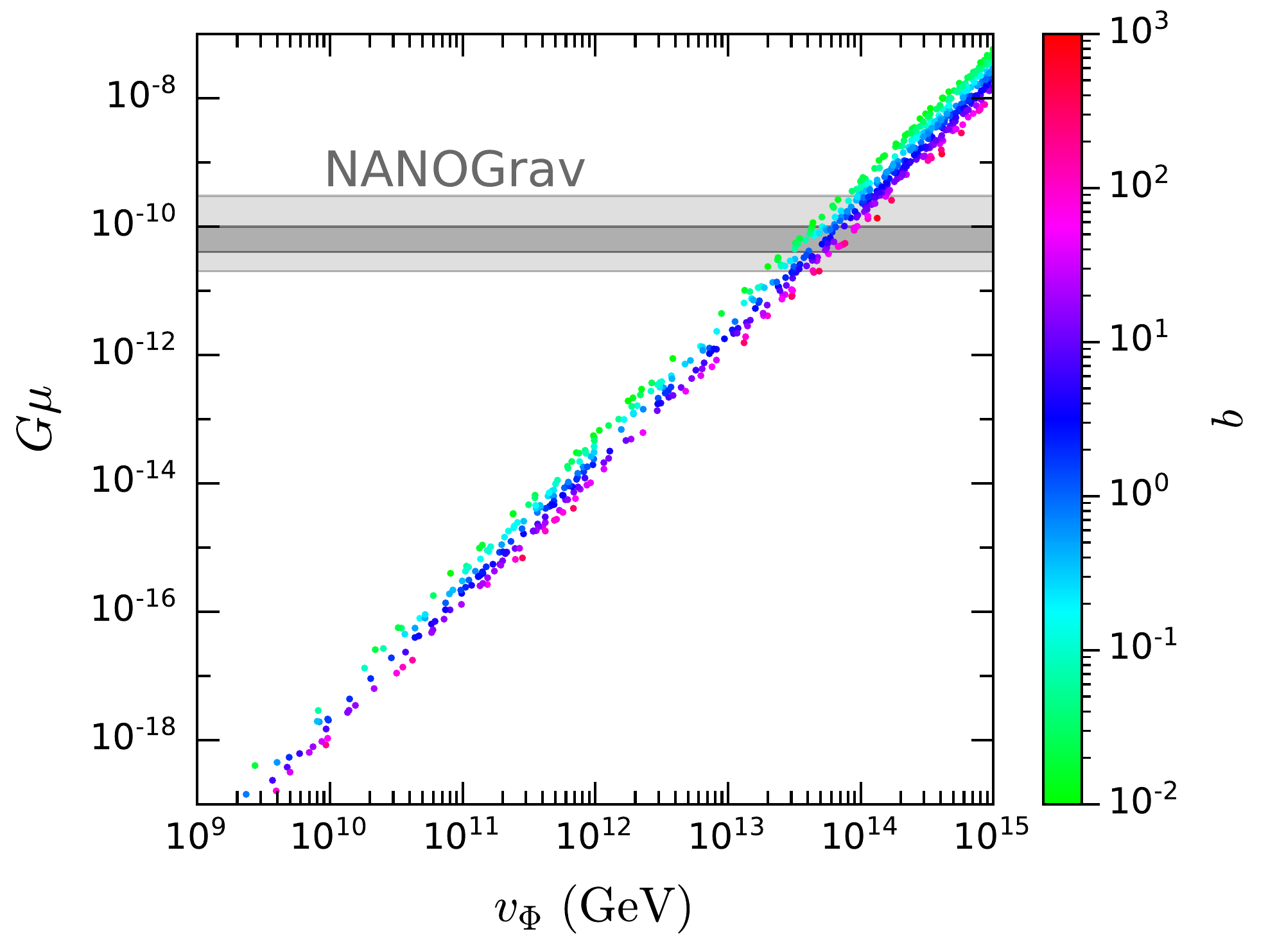}
\caption{68\% (gray) and 95\% (light gray) confidence intervals of $G\mu$ for the NANOGrav anomaly compared with the selected parameter points in the pNGB DM model.}
\label{fig:suspected_signal}
\end{figure}

Finally, we discuss a related experimental anomaly.
According to the 12.5-yr data set, the NANOGrav collaboration reported strong evidence of a stochastic common-spectrum process, but they found no significant evidence of the quadrupolar spatial correlations, which are necessary to claim the detection of an SGWB~\cite{NANOGrav:2020bcs}.
If this signal is genuine, it can be explained by the SGWB induced by cosmic strings, and the 68\% (95\%) confidence interval of the string tension implied by the data is $\num{4e-11} < G\mu < 10^{-10}$ ($\num{2e-11} < G\mu < \num{3e-10}$)~\cite{Ellis:2020ena}.
In Fig.~\ref{fig:suspected_signal}, we compare the NANOGrav confidence intervals with the selected parameter points on the $v_\Phi$-$G\mu$ plane and find that it is possible to explain such a suspicious signal at 95\% C.L. by $\num{2e13}~\si{GeV} \lesssim v_\Phi \lesssim \num{2e14}~\si{GeV}$ in the pNGB DM model.
Of course, more data from future GW experiments are needed to reveal the nature of this anomaly.

\section{Summary and outlook}
\label{sec:sum}

We studied the stochastic GW signals from cosmic strings induced by the spontaneous breaking of the hidden $\UoneX$ gauge symmetry from the UV-complete model for pNGB DM.
Because of the pNGB nature of the DM candidate $\chi$, the tree-level $\chi$-nucleon scattering cross section is highly suppressed by the UV scale $v_\Phi$, which characterizes the breaking scale of the $\UoneX$ gauge symmetry.
Thus, DM direct detection experiments would not be able to probe $\chi$ for $v_\Phi \gtrsim 10^5~\si{GeV}$.
Meanwhile, $\chi\chi$ annihilation processes are not suppressed, and the observed DM relic abundance can be easily achieved by the conventional freeze-out mechanism.
Therefore, the pNGB DM model can naturally explain dark matter in the universe and satisfy current experimental constraints.

In order to investigate the phenomenological constraints on the pNGB DM model, we carried out a random scan in the 10-dimensional parameter space.
The parameter points simultaneously satisfying the constraints from the DM lifetime, the DM relic abundance, indirect detection experiments, measurements of the SM-like Higgs boson $h_1$, and collider searches for the exotic Higgs boson $h_2$ have been selected.

The spontaneous $\UoneX$ symmetry breaking at the UV scale $v_\Phi$ is expected to generate cosmic strings in the early universe.
The intersections among the cosmic strings would lead to closed loops, which could emit GWs via relativistic oscillations.
The incoherent superposition of GWs leads to an SGWB, which is an important target for GW experiments.
We evaluated the corresponding GW spectra for the viable parameter points according to the BOS and LRS models for loop production.

Moreover, we studied the constraints from current GW experiments including NANOGrav, PPTA, and LIGO-Virgo.
These constraints excluded the parameter points with $v_\Phi \gtrsim \num{5e13}$ $(\num{7e11})~\si{GeV}$ for the BOS (LRS) model.
Furthermore, we estimated the sensitivity of future GW experiments LISA, Taiji, TianQin, CE, and SKA.
For the BOS (LRS) model, these future experiments could explore the parameter points with $v_\Phi$ down to $\sim\num{2e10}~(\num{5e9})~\si{GeV}$.
Note that the bound on the DM lifetime has basically excluded a UV scale $v_\Phi$ lower than $10^9~\si{GeV}$.
Therefore, almost all the viable parameter points of the pNGB DM model can be well studied in the future.

Typical experiments in particle physics would lose their sensitivity if the related energy scale is too high.
In this study, however, a higher UV scale $v_\Phi$ would lead to a higher tension of cosmic strings, resulting in an SGWB with higher energy density, which would be more easily discovered in GW experiments.
Remarkably, we have demonstrated that GW experiments could be complementary to other types of experiments in exploring new physics beyond the SM.

In the calculation above, the value of $v_{\Phi}$ is assumed to be a constant.
Nonetheless, if the temperature corrections to the effective potential of the scalar fields are considered, $v_{\Phi}$ would depend on the temperature at the epoch when the spontaneous breaking of the $\UoneX$ gauge symmetry occurs.
This could affect the tension of the cosmic strings formed and hence the SGWB spectrum.
The treatment in this paper should be regarded as an approximation without the temperature effect.
Such a effect is worthy to be studied in the future.

\begin{acknowledgments}

We thank Shi-Qi Ling for helpful discussions.
This work is supported by the National Natural Science Foundation of China under Grant No.~11805288.

\end{acknowledgments}

\bibliographystyle{utphys}
\bibliography{ref}
\end{document}